\documentclass[
	amsmath,
	amssymb,
    aps,
    reprint,
    pre,
    showkeys,
]{revtex4-1}

\usepackage{array}
\usepackage[utf8]{inputenc}
\usepackage{amsthm}
\usepackage{xcolor}
\usepackage{hyperref}
\usepackage{graphicx}
\usepackage[font=small,justification=centerlast,format=plain]{caption}
\usepackage{subcaption}
\usepackage{xcolor}
\usepackage{dsfont}
\usepackage{booktabs}
\usepackage{braket}
\usepackage{bm}
\usepackage[capitalise]{cleveref}
\usepackage{tikz}
\usepackage{pgf} 
\usepackage{color}

\usepackage{amsthm}

\theoremstyle{plain}
\theoremstyle{definition}
\usepackage{afterpage}

\begin{document}
\title{SHEEP: Signed Hamiltonian Eigenvector Embedding for Proximity}

\author{Shazia'Ayn Babul}
\email{shazia.babul@maths.ox.ac.uk}
\affiliation{Mathematical Institute, University of Oxford, Oxford, UK}
\author{Renaud Lambiotte}
\email{renaud.lambiotte@maths.ox.ac.uk}
\affiliation{Mathematical Institute, University of Oxford, Oxford, UK}
\affiliation{Turing Institute, London, UK}

\date{\today}

\begin{abstract}
We introduce a spectral embedding algorithm for finding proximal relationships between nodes in signed graphs, where edges can take either positive or negative weights. Adopting a physical perspective, we construct a Hamiltonian which is dependent on the distance between nodes, such that relative embedding distance results in a similarity metric between nodes.  The Hamiltonian admits a global minimum energy configuration, which can be re-configured as an eigenvector problem, and therefore is computationally efficient to compute. We use matrix perturbation theory to show that the embedding generates a ground state energy, which can be used as a statistical test for the presence of strong balance, and to develop an energy-based approach for locating the optimal embedding dimension.  Finally, we show through a series of experiments on synthetic and empirical networks, that the resulting position in the embedding can be used to recover certain continuous node attributes, and that the distance to the origin in the optimal embedding gives a measure of node “extremism”.  
\end{abstract}


\maketitle

\section{Introduction}
Networks are powerful representation of complex systems, comprising of a collection of nodes joined by edges that represent interactions~\cite{newman2010networks}. Typical examples stem from an array of disciplines, including sociology (social contact networks) urban infrastructures (public transportation networks), or biological interactions (gene interaction networks).  Edges represent diverse types of interactions, which can be distinguished by attributes on the edges, such as edge weights, to identify the nature and intensity of the relationship.  In this paper, we consider \textit{signed networks}, where the edges can have either positive or negative weight. A fundamental example is a social network, where positive edges represent friendship and negative edges represent animosity between individuals. In what follows, we adopt the convention for our figures that red denotes a negative edge, while blue is a positive edge. 

\begin{figure}
\centering
\includegraphics[scale=0.2]{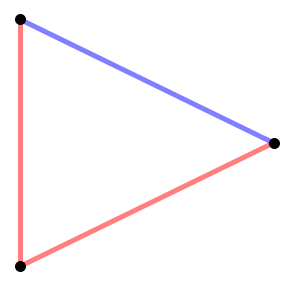}
\includegraphics[scale=0.2]{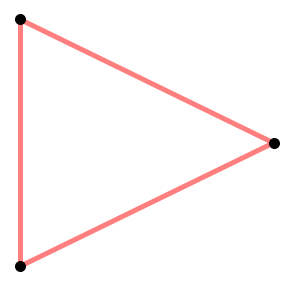}
\includegraphics[scale=0.2]{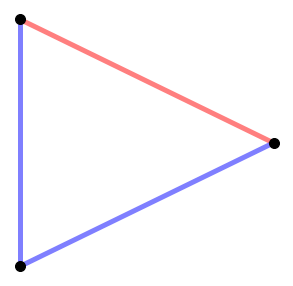}
\caption{Strongly balanced cycles (left) have even numbers of negative edges. Weakly balanced cycles (middle) have $i$ negative edges, where $i \neq 1$. Unbalanced cycles (right) have a single negative edge.}
\label{fig:balance_ex_cycle}
\end{figure}

\begin{figure}
\centering
\includegraphics[angle = 90, scale=0.25]{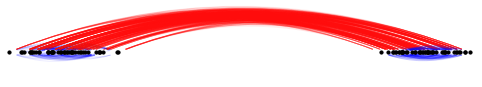}
\includegraphics[scale=0.25]{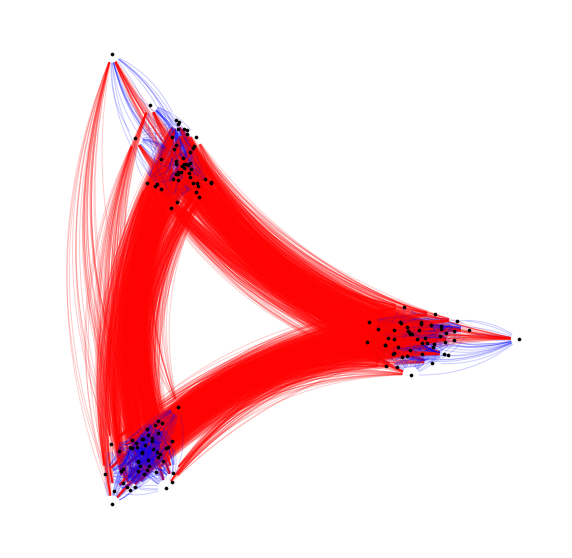}
\caption{Strongly balanced signed networks (left) have two opposing factions with positive edges inside and negative edges between. Weakly balanced networks (right) can have 3 or more factions.}
\label{fig:balance_ex}
\end{figure}

The topology of signed social networks is typically understood through Heider’s theory of structural balance, a psychological theory which argues that “balanced” structures are more favourable in interpersonal relationships, following the adage \textit{an enemy of my enemy is my friend} and that a friend of a friend is also a friend~ ~\cite{heider1946attitudes}. Building this notion into the framework of signed networks, Harary argued in  \cite{harary1953notion} that certain triadic patterns of signed relationships exhibit “balance” and therefore should be more common in social networks. In \cite{cartwright1956structural}, Cartwright and Harary use the term “balance” to refer to a signed network where all cycles contain an even number of negative edges, as in Figure \ref{fig:balance_ex_cycle}. This condition is now commonly referred to as \textit{strong balance}. Social balance is central to understanding the multi-scale topology of signed networks due to the theorem of Harary which equates strong balance with \textit{bi-polarization}; a strongly balanced graph is clusterable, meaning the nodes can divided into two groups with positive edges inside, and negative edges connecting them~\cite{harary1953notion}. A weaker condition for clusterability was proved by Davis in \cite{davis1967clustering}, using the notion of \textit{weak balance} to refer to graphs where no cycle has only a single negative edge. Graphs which exhibit weak balance, can be partitioned into $k$ clusters with positive edges inside, and negative edges connecting them. Identifying the optimal partition of a signed graph into $k$ clusters, where $k$ is unknown, has been shown to be NP hard and many different methods have been proposed to solve this problem numerically~\cite{aref2021identifying, traag2009community, doreian2009partitioning}. Figure \ref{fig:balance_ex_cycle} shows examples strong balanced, weak balanced and unbalanced cycles, while Figure \ref{fig:balance_ex} depicts a strongly balanced (2 faction) graph and a weakly balanced (3 faction) graph. These conditions are deeply related to the Ising spin glass model, where negative plaquettes introduce geometric frustration, analogous to the unbalanced k-cycles on signed graphs~\cite{cao2018frustration}. We give an overview of the related literature in the following section, and direct the reader to \cite{10.1145/2956185} for a more detailed overview of signed network properties and principles.

Structural balance provides a valuable framework for understanding signed networks, relating cycle patterns to graph clusterability. As a consequence, the relationship between two nodes is related not just to the (local) sign of the link between them, but to the whole global topology. Often, the cluster structure of signed networks is investigated through network embedding. Network embedding refers to a method for obtaining a low-dimensional representation of nodes, taking into account multi-scale graph topology. In general, embedding methods have a range of applications from clustering to semi-supervised node classification, and are used in recommender systems to make recommendations based on proximity in a latent (embedding) space~\cite{Zhang_2021,fouss2016algorithms}. For signed networks, embedding methods typically seek to cluster nodes in the same faction close together in space, while simultaneously separating each faction. Yet, it has been argued in the literature that empirical networks are often not perfectly balanced, and the tendency for signed social networks to become balanced may not even hold over time~\cite{doreian2001pre}. In particular, signed networks may not have a natural cluster structure, and as proposed in \cite{doreian2009partitioning}, the formation of unbalanced signed social networks may be influenced by a combination of processes including mediation, differential popularity and internal subgroup hostility, resulting in different patterns than structure balance alone.  When ground truth clusters do not exist, it may be more useful to quantify the similarity between nodes in terms of a continuous distance variable, eliminating the need to make assumptions about faction numbers and memberships. This is particularly true when "mediators" exist in the graph, a role that refers to nodes with positive edges to members of many different factions~\cite{doreian2009partitioning}. 

In this article, we adopt a physical viewpoint to define SHEEP, an energy-based signed network embedding method, incorporating local, intermediate and global information into a proximity measure between nodes. Modeling the nodes as a system of particles interacting with attractive forces (positive edges) and repulsive forces (negative edges), we construct a Hamiltonian which is dependant on the distance between nodes, such that relative embedding distance results in a similarity metric. Unlike embedding methods which focus on partitioning, and are useful for binary prediction, we show through a series of experiments on synthetic and empirical networks, that the resulting position in the embedding can be used to recover certain continuous node attributes.  In addition, the Hamiltonian admits a global minimum energy configuration, which can be re-configured as an eigenvector problem, and therefore is computationally efficient to compute. At the global level, the embedding generated by the minimum energy configuration is intrinsically related to structural balance, and the ground state energy can be used as a statistical test for bi-polarization.

This article is organized as follows. In Section II, we give an overview of the related literature on network embedding methods, with a particular focus on spectral and energy-based approaches. In Section III, we present our Hamiltonian and give the motivation for the spectral solution. Section IV focuses on the case of strong balance. We present three main results on the relationship between the ground state energy and the balance of the graph, proposing a bi-polarization measure. Section V extends our method to higher dimensions, illustrating an energy-based approach to finding the optimal embedding dimension. Finally, in Section VI, we propose two applications of the signed network embedding method; (1) a way to recover continuous node attributes based on their proximity in the embedding, and (2) using the node's distance to the origin as a measure of "extremism", evaluating the method on both synthetic generated signed networks and empirical networks. 

\subsection{Notation}
For a matrix $M \in \mathbb{R}^{n x n}$ we denote the eigenvalues $\lambda_{i}(M)$ with associated eigenvectors $\nu_{i}(M)$. A symmetric matrix $M$ has an eigenvalue ordering $\lambda_{1}(M) \geq \lambda_{2}(M) ... \geq \lambda_{n}(M)$ with associated eigenvectors. We denote $\textbf{1}$ as the all ones vector. The graph $G = (V, E)$ is the signed graph with node set $V$ and edges $E$, with $E^{+}$ being the set of positive signed edges, and $E^{-}$ being the set of negative signed edges, such that $E^{+} \cup E^{-} = E$ and $E^{+} \cap E^{-} = \emptyset$. 

If $A$ is the adjacency matrix of graph $G$, then $A_{ii} = 0, \forall i \in V$. In what follows, we focus on unweighted signed networks. For an edge $(i,j) \in E^{+}$, it follows that $A_{ij} = 1$ and likewise, an edge $(i,j) \in E^{-}$ corresponds to the matrix entry $A_{ij} = -1$. The positive adjacency matrix $A^{+}$ is defined by $A^{+}_{ij} = 1$ if $A_{ij} = 1$, and 0 elsewhere. Similarly, the negative adjacency matrix $A^{-}_{ij}$ is defined by $A^{-}_{ij} = -1$ if $A_{ij} = -1$, and 0 elsewhere, such that $A = A^{+} + A^{-}$. We also have that $D^{+} = \text{diag}(deg_{1}^{+}, deg_{2}^{+}...deg_{n}^{+})$, where $deg^{+}_i = \sum_j A_{ij}^+$, the degree matrix of the positive sub-graph, and $D^{-} = \text{diag}(deg_{1}^{-}, deg_{2}^{-}...deg_{n}^{-})$, where $deg^{-}_i = - \sum_j A_{ij}^-$,  the degree matrix of the negative sub-graph, where in particular we note that $deg_{i}^{-} \geq 0$. We also use the convention that an n-complete graph is a complete graph with $n$ nodes, such that every possible edge exists. 

The literature does not have a consistent naming system for the different Laplacians we refer to here. What we call the opposing Laplacian after \cite{shi2019dynamics} is also known as the signed Laplacian as in \cite{kunegis2010spectral}. Following \cite{shi2019dynamics} we define the repelling Laplacian, which is also known as the physics Laplacian in \cite{fox2017numerical}, the net Laplacian in \cite{ou2021net}, or simply the unsigned Laplacian due to its identical construction to the Laplacian on unsigned graphs.

\section{Related Literature}
\subsection{Spectral Methods}
Classically, network embedding methods fall into two categories; spectral methods and physical methods, which are useful for different purposes.  While spectral methods on signed networks are primarily designed for graph clustering, which is not our objective here, it is nevertheless a rich field and we give an overview of the relevant literature as follows. Spectral clustering on signed graphs began with \cite{anchuri2012communities}, where the authors propose a clustering method by minimizing an objective function associated with a modularity matrix.  In \cite{kunegis2010spectral}, Kunegis et al. propose a spectral method using the opposing Laplacian defined $L_{o} = \bar{D} - A$ where $\bar{D}$ is the diagonal matrix associated to the unsigned version of the graph where the degrees are obtained degrees as $\sum_j |A_{ij}|$. They embed the signed graphs in two dimensions, using the first two eigenvectors of the opposing Laplacian. The resulting inner product arising from the opposing Laplacian is not a function of distance in the embedding, and the presence of many or few negative links between positively connected factions does not effect the final positions, unlike the physically inspired method that we propose here. In particular, the opposing Laplacian identifies whether the graph is strongly balanced or not, but does not given an indication of the intensity of the negative interactions between the nodes.  Extending the opposing Laplacian method to higher dimensions was shown to have significant weaknesses in \cite{chiang2012scalable}, where the authors propose a modified method by minimizing an objective called the Balanced Normalized Cut. Relatedly, in \cite{xiao2020searching}, the authors use a normalized version of this opposing Laplacian as a basis for finding a locally optimal polarized communities, given a set of input nodes. In \cite{cucuringu2019sponge}, Curcuringu et al. construct a regularized spectral algorithm that performs particularly well at recovering clusters, even when graphs are sparse.

Most similar to our work here is the work of \cite{knyazev2018spectral} in which Knyazev uses the repelling Laplacian $L_{r} = D - A$ where $D$ is the diagonal matrix of row sums of the adjacency matrix $A$, which can take positive and negative values. Knyazev argues that signed graphs can be clustered effectively using the repelling Laplacian from a mechanical perspective, modeling clusters as eigenmodes of a mass-spring system with negative springs. The graph is clustered using the $k$ smallest eigevectors, where $k$ is chosen by looking for the largest eigen-gap. Although we propose to use the same Laplacian here, we work from an energy based approach, seeking proximal relations instead of clusters, and demonstrating that the ground state energy is a statistical test for polarization. In \cite{fox2017numerical}, the authors apply QR factorization on the eigenvalues of Gremban's expansion to obtain clusters. Gremban's expansion is a matrix related to a random walk process on a signed network~\cite{tian2022spreading}. \cite{fox2018investigation} provides a short review of related spectral clustering methods used for signed graphs, comparing the opposing Laplacian, the repelling Laplacian, and Gremban's expansion.  An approach proposed in \cite{bonchi2019discovering} searches for a bi-partition of a two-way polarized network allowing for "neutral" actors, by maximizing an objective function associated with the quadratic form of the adjacency matrix. While this approach makes use of spectral methods, the resulting optimization problem is NP-hard, and instead obtains approximate solutions. This method has been extended to $k$-clusters in \cite{tzeng2020discovering}. As pointed out in \cite{cucuringu2019sponge}, most spectral embedding methods fix \textit{a priori} the number of clusters $k+1$, embed the nodes using the $k$ smallest eigenvectors, and finally apply a k-means analysis to identify clusters. In our proposed method, we use an energy minimization method to select the optimal embedding dimension. 

\subsection{Physical Methods}
Energy-based techniques for studying signed networks have been proposed predominantly to detect clusters as in \cite{traag2009community, he2020energy}, where the proposed Hamiltonian is a function of the node cluster assignments, and the sign of the edge between them. These methods draw from spin glass literature, using simulated annealing to locate the optimal cluster assignment. For unsigned networks, however, continuous energy-based techniques are another prominent family of graph embedding methods. Typically, an energy function is chosen by assigning attractive and repulsive forces to edges, and an optimization algorithm is employed to find a minimum energy configuration.  For unsigned graphs, edges are often modeled as attractive forces, while all nodes repulse one another to achieve the best-looking two-dimensional layout. The Spring Embedder method introduced in \cite{eades84} was one of the earliest force directed models, treating nodes as electric charges with springs modeling edges. The algorithm used to locate the minimum energy configuration was further refined by Fruchterman and Reingold in \cite{fruchterman1991graph}. In \cite{kamada1989algorithm}, Kamada and Kawai introduce an algorithm which treats edges as springs, with rest length equal to the graph theoretical distance. In \cite{kermarrec2011energy}, Kermarrec et al. introduce an energy based approach for drawing signed graphs in two dimensions to detect signed clusters, using a logarithmic energy function as proposed in \cite{noack2007energy} for finding communities in unsigned graphs. 

Unless the chosen energy function is strictly convex, the optimization algorithms employed for force-based methods locate a local minima, which is acceptable for the purposes of graph drawing. Here, we instead construct a Hamiltonian which admits a global minima. We draw inspiration from SpringRank, a physically inspired model that generates hierarchical rankings of nodes in directed, unsigned networks along a one dimensional line, using a spring potential based Hamiltonian~\cite{de2018physical}. The SpringRank Hamiltonian is given by: 
\begin{equation}
    H(\textbf{s}) = \frac{1}{2} \sum_{i,j}A_{ij}(s_{i} - s_{j} - 1)^{2}
\end{equation}
The SpringRank Hamiltoninan has a global minima which allows for statistical comparisons of the ground state energy, to determine significance of the hierarchical structure. In what follows, we generalize the SpringRank method to undirected signed networks, embedding the nodes in a k-dimensional space by defining a Hamiltonian that is the sum of repulsive and attractive forces corresponding to negative and positive edges respectively, and finding the global minimum energy configuration. 

\section{Spring/Anti-Spring Signed Graph Hamiltonian}
\subsection{Hamiltonian}
The generalization of the spring forced based Hamiltonian on a signed network is constructed as follows. Suppose each node $i$ has an associated position vector $\textbf{x}_{i}$ in metric space $\mathbb R^{k}$. Positive edges are modeled as spring attractive forces, and negative edges are associated with an "anti-spring" repulsive force, which is similarly quadratic in distance. 
\begin{equation}
\label{springham}
    H = \sum_{i,j}A^{+}_{ij} |\textbf{x}_{i} - \textbf{x}_{j}|^{2} + \sum_{i,j}A^{-}_{ij} |\textbf{x}_{i} - \textbf{x}_{j}|^{2}
\end{equation}
The first term of the Hamiltonian is minimized when positively connected nodes have minimal distance between them, while the second term requires negatively connected nodes to have the largest possible distance between them. We seek to find the set of position vectors $\{\textbf{x}_{i}\}$ that describe the minimum energy configuration. When a negative edge exist between two nodes, their interaction term may be minimized when the two nodes are pushed apart to infinity. Consequently, we introduce a constraint on Eq. \ref{springham} to control for the "explosion" of distance. 

\subsection{Relation to Spectral Methods}

Before introducing the constraint, we first illustrate the relationship between the proposed energy based embedding, and spectral approaches. As graph with strong balance can be separated into two factions, we constrain ourselves to a one dimensional embedding for the remainder of this section. The vector $\textbf{x}$ denotes the $n$-dimensional vector describing the positions $x_i$ of the $n$ nodes along a line.  

In \cite{shi2019dynamics}, the authors introduce two types of signed graph Laplacian. The Laplacian of the positive part of the graph is $L^{+} = D^{+} - A^{+}$. There are two possibilities for the negative graph Laplacian, corresponding to two possibilities for the signed Laplacian. The opposing Laplacian: 
\begin{equation}
    L_{o} = D^{+} - A^{+} + D^{-} - A^{-}
\end{equation}
And the repelling Laplacian: 
\begin{equation}
    L_{r} = D^{+} - A^{+} - D^{-} - A^{-}
\end{equation}
These two matrices induce quadratic forms on the vector $\textbf{x} \in \mathbb{R}^{n}$. Note that $x_{i}$ is a scalar value associated with the node $i$. The induced quadratic form of the opposing Laplacian is:
\begin{equation}
   \textbf{x}^{T} L_{o} \textbf{x} = \sum_{i,j}A^{+}_{ij} |x_{i} - x_{j}|^{2} - \sum_{i,j}A^{-}_{ij} |x_{i} + x_{j}|^{2}
\end{equation}
And the repelling Laplacian: 
\begin{equation}
    \textbf{x}^{T} L_{r} \textbf{x} = \sum_{i,j}A^{+}_{ij} |x_{i} - x_{j}|^{2} + \sum_{i,j}A^{-}_{ij} |x_{i} - x_{j}|^{2}
\end{equation}
The quadratic form resulting from the repelling Laplacian is equivalent to the spring-inspired Hamiltonian proposed in Eq. \ref{springham}, for one-dimensional position vectors. The repelling Laplacian is symmetric and real, so its eigenvectors can be chosen to be ortho-normal. Since this matrix has zero row sum, it has a 0 eigenvalue associated to the eigenvector $\textbf{1}$. Unlike the opposing Laplacian, which has been previously used for spectral embedding as in \cite{kunegis2010spectral} because it is positive semi-definite, the repelling Laplacian is indefinite. From a physical perspective, it describes a Hamiltonian that permits "explosions" of  distances due to the quadratically increasing energy of repulsive forces. 

\subsection{Spherical Constraint}

Following a method proposed in \cite{kawamoto2022consistency} for ordering nodes in graphs, we impose the spherical constraint condition that $\sum x_{i}^{2} = R$. This corresponds to the minimization of the quadratic form associated to the repelling Laplacian, with a Lagrange multiplier constraint, or:
\begin{equation}
\textbf{x}^{T}L_{r}\textbf{x} - \beta (\sum x_{i}^{2} - R)  
\end{equation}
Taking the gradient results in following equation for eigenvector $\nu$ associated with eigenvalue $\lambda$: 
\begin{equation}
    L_{r} \nu = \lambda \nu
\end{equation}
The smallest eigenvalue is associate with eigenvector $\nu_{n}$, which is normalized such that $\nu^{T}_{n} \nu_{n} = R$. The minimum value of the Hamiltonian is then: 
\begin{equation}
\label{replapmin}
    H(\nu_{n}) = \nu^{T}_{n}L_{r}\nu_{n} = \lambda_{n} R
\end{equation}
The spherical constraint is itself non-convex, but a symmetric matrix like the repelling Laplacian can be diagonalized as:
\begin{equation}
\label{orthodecomp}
    L_{r} = UDU^{-1}
\end{equation}
Where $U$ is an orthogonal matrix which maps $\textbf{x}$ to another point on the unit $n$-sphere. It follows that the global minimum of the Hamiltonian is given by the minimal eigenvalue of Eq. \ref{replapmin}. In what follows we take $R = 1$. Note that unlike the other force based approaches which require an optimization procedure, finding the minimal value of the Hamiltonian with the spherical constraint reduces to an eigenvalue problem, which far more computationally efficient.

We note here that a related approach using proximity instead of partitioning was taken in \cite{brandes2006summarizing, brandes2008visual}, where the authors describe a two-dimensional graph drawing technique for signed networks onto a "conflict space" using the smallest eigenvalues of the adjacency matrix. The advantage of using the repelling Laplacian comes from the physicality of the system as a consequence of Eq. \ref{springham}, and the resulting energy associated with the network. We also show how our method can be generalized to higher dimensions, which is the key distinction between a graph \textit{drawing} and graph \textit{embedding} method. 

\section{Ground State Energy and Strong Balance}
In this section, we propose a test statistic for bi-polarization, by exploiting the physicality of our system; the ground state energy, denoted as $E^{0}$. Analogously to the Ising model, where the lowest ground state energy is achieved when there is no geometric frustration introduced by negative cycles, we want to show that the ground state energy of our Hamiltonian is minimized when the graph exhibits strong balance. If a test statistic is significant, it should be highly improbable on a null model. As the ground state energy depends on many aspects of the network structure, we focus on the following null model: fixing the graph topology, we randomly reshuffle the signs of the edges, while preserving the density of positive and negative edges, as in \cite{leskovec2010signed, szell2010multirelational}. If the real network has a lower ground state energy then the networks produced by the null model, we can conclude that the network has significant underlying bi-polarization structure.  Since the ground state energy of our Hamiltonian is the minimum eigenvalue of the repelling Laplacian, we present here some new results on the spectrum of the repelling Laplacian for complete-graphs, and then generalize the results numerically to non-complete graphs.

\subsection{Repelling Laplacian: Spectral Results}
We consider the set of complete signed graphs with $n$ nodes, subject to the condition that $E^{+}, E^{-} \neq \emptyset$, denoted by $\{G^{n}\}$. We present three results: (1) The minimal eigenvalue $\lambda_{n}$ is bounded from below by $-n$. (2) The lower bound $\lambda_{n} =  -n$ is achieved when the graph has a perfect bi-partition (strong balance). And (3) Introducing frustration by switching edge signs will strictly increase the value of the ground state energy. These results motivate the use of the ground state energy as a bi-polarization measure, or measure of strong balance. 
\\ \\
\textbf{Theorem 1.} 
\emph{For an n-complete signed graph subject to the condition that $E^{+}, E^{-} \neq \emptyset$, the smallest eigenvalue of the repelling Laplacian $\lambda_{n} \geq -n$.}
\\ \\
\normalfont The proof is deferred to the appendix, but the main steps involved are as follows. Beginning with the n-complete graph of all negative edges, one can easily show that the spectrum is $-n$ with multiplicity $n-1$ and 0 with multiplicity 1. Any other signed $n$-complete graph can be reached by successively changing an edge sign to positive, associated with the addition of a $L^{shift}$ to the repelling Laplacian, which is positive semi-definite. Applying Weyl's inequality to the sum of the two Laplacians gives the required result.  

Next, we show that the lower bound is reached when the graph is strongly balanced and admits a bi-partition. 
\\ \\
\textbf{Theorem 2.}  \emph{Consider a strongly balanced graph $\tilde{G} \in \{G^{n}\}$, such that the nodes of graph $\tilde{G}$ can be partitioned into two sets $V_{1}$ and $V_{2}$, where $V_{1}, V_{2} \neq \emptyset$. Nodes inside each set are connected with positive edges, while the edges connecting the sets are negative. The smallest eigenvalue of the repelling Laplacian is $\lambda_{n} = -n.$} 
\\ \\

\normalfont Again, we defer the proof to the Appendix. Broadly, we construct the eigenvector $\nu$ associated to the minimum eigenvalue $ \lambda_{n}$ by placing nodes in the same set at the same point such that for node $k \in V_{i}$ we set $\nu_{k} = x_{i}$ where $i \in (1,2)$. Using the orthonormality of the eigevectors, we can show that associated eigenvalue is $-n$. 
\begin{equation}
\label{gse}
    E^{0}_{bal} = \lambda_{n}= -n
\end{equation}
By Theorem 1, this is the minimum eigenvalue. 
By Theorems 1 and 2, the ground state energy reaches the lower bound when the graph is non-frustrated, and has a perfect bi-partition in accordance with strong balance.  When the graph is "frustrated" we want to show that the ground state energy, $E^{0}_{frus}$ is larger than the energy associated with the balanced graph $E^{0}_{bal}$ given in Eq. \ref{gse} as $-n$. We compare between graphs that have the same number of nodes, $n$, as well as the same density of positive and negative edges.
\\ \\
\textbf{Theorem 3.}  \emph{Consider a strongly balanced graph $\tilde{G} \in \{G^{n}\}$, such that the nodes of graph $\tilde{G}$ can be partitioned into two sets $V_{1}$ and $V_{2}$, where $V_{1}, V_{2} \neq \emptyset$. Introduce frustration into the graph by switching two edge signs. When the graph is sufficiently large such that $n > 4$, and $V_{1}, V_{2} > 2$, the ground state energy is strictly increased $E^{0}_{frus} > E^{0}_{bal}$.}
\\ \\
The proof involves writing the repelling Laplacian of the frustrated graph as the sum of the repelling Laplacian of the balanced graph $L^{\tilde{G}}_{r}$, and a perturbation, $L^{switch}$, associated to the switching of a positive and negative edge sign.  The new graph is "frustrated" because the two terms $L^{\tilde{G}}_{r}$ and $L^{switch}$ cannot be simultaneously minimized. Finding a bound on the second smallest eigenvalue of the repelling Laplacian of the balanced graph $L^{\tilde{G}}_{r}$, gives an approximation for the "energy gap" between the ground and first energy levels. When the graph is sufficiently large such that $n > 4$, and $V_{1}, V_{2} > 2$, we can use the energy gap to bound the ground state energy from below, obtaining the inequality we require. In practice, $n > 4$ is a weak constraint, as most graphs of interest contain more than four nodes.
\begin{equation}
E^{0}_{frus} > -n  = E^{0}_{bal}
\end{equation}

\subsection{Bi-Polarization Measure}
Following the intuition provided by the spectrum of the repelling Laplacian on complete signed graphs, we numerically generalize to the case of non-complete graphs, showing that the ground state energy is a statistically significant measure of bi-polarization, as compared to the null model. The test correctly identifies the polarization structures on various synthetic signed graphs with strong balance. We propose a bi-polarization score using the z-score of the ground state energy of the graph, compared to the the null model energy distribution. We define:
\begin{equation}
    Z(G) = \frac{E^{G}_{0} - \braket{E_{0}}}{\sigma}
\end{equation}
Where $\braket{E_{0}}$ is the mean of the null model distribution, and $\sigma$ is the standard deviation. A large negative value of $Z(G)$ indicates that the graph is significantly polarized. 

There exist a number of proposed bi-polarization measures for signed networks, which typically focus on the local or global properties of the graph. For example,   \cite{Kirkley_2019} proposes a balance measure that counts the number of unbalanced cycles in the graph, inversely weighted by length.  A more global measure is the frustration index, which counts the number of frustrated edges associated with the best bi-partition \cite{aref2019balance}. An intermediate scale measure, POLE, takes the correlation between a node's signed and unsigned random walk, based on the assumption that link sign should be related to unsigned community structure~\cite{huang2021pole}. Our measure incorporates both local and global structure, and allows for comparison between networks with fixed size and edge density.   

\begin{figure}
\centering
\includegraphics[scale=0.44]{sbm100.png}
\includegraphics[scale=0.34]{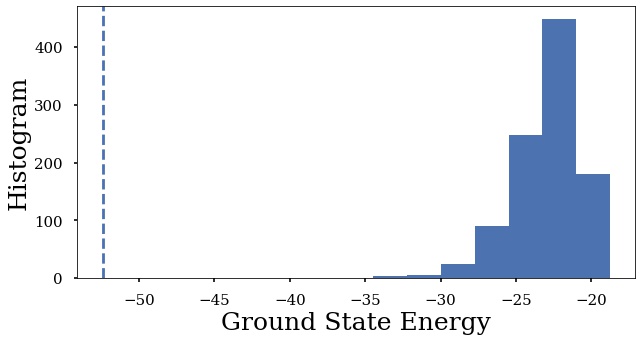}
\caption{2-community SSBM embedded in one-dimensions using SHEEP (top). Histogram of the ground state energies of the 1000 realization of the null model, with the SSBM ground state energy shown with the dotted line (bottom). }
\label{fig:sbmzscore100}
\end{figure}

\begin{figure}
\centering
\includegraphics[scale=0.44]{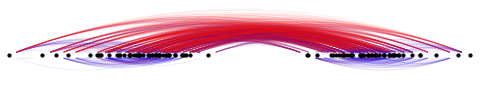}
\includegraphics[scale=0.34]{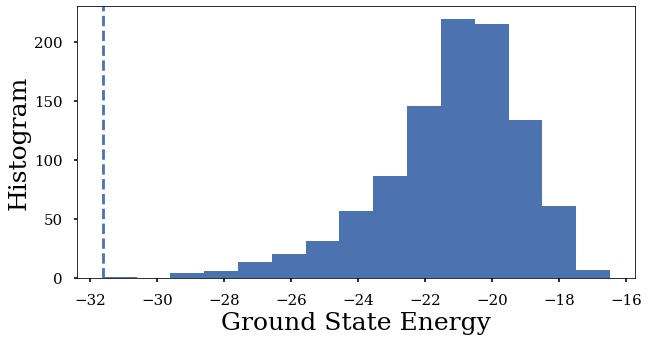}
\caption{2-community SSBM with 0.2 probability of edge sign flips, embedded in one-dimensions using SHEEP (top). Histogram of the ground state energies of the 1000 realization of the null model, with the SSBM ground state energy shown with the dotted line (bottom).}
\label{fig:sbmzscore80}
\end{figure}

We numerically test the bi-polarization measure on synthetic signed networks with polarization structure. A signed stochastic block model (SSBM) is a synthetic graph of $n$ communities with positive edges inside the communities, and negative edges between. Here, we show the results on 2-community SSBMs of equal community sizes [50,50]. For each pair of nodes, the link probability (positive or negative) is 0.5. Because such graphs trivially have a bi-partition, they also obey strong balance. Figure \ref{fig:sbmzscore100} shows the SHEEP (repelling Laplacian first eigenvector) embedding of the SSBM, red indicating negative edges and blue indicating positive. The histogram compares the ground state energy of the balanced SSBM to 1000 realizations of the null model. The resulting bi-polarization measure $Z(G_{bal})$ is -12.24, correctly identifying highly significant bi-polarization structure. Next, we introduce noise into the SSBM by flipping each edge sign with probability 0.2. Figure \ref{fig:sbmzscore80} shows the ground state energy and null model distribution for the noisy graph. For the initial realization of the SSBM, the $Z(G_{0.8})$ score is -4.87, which is still significant. As the two SSBMs in Figures \ref{fig:sbmzscore100} and \ref{fig:sbmzscore80} have the same number of nodes and edges, the bi-polarization scores can be compared to correctly conclude that the non-noisy one is more polarized. 

\section{Weak Balance and Extension to Higher Dimensions}

In one dimension, we can understand graphs in terms of strong balance, because the bi-partition can be projected along the one dimensional line. Weakly balanced graphs are $k$-clusterable, meaning they can be partitioned into $k$ factions. This also implies that representing the relationships between nodes may not be possible using a one dimensional projection. For noisy and un-balanced graphs, this is even more important; more dimensions may be required to ensure that the distance between nodes in the embedding represents a proximity measure that takes into account the global graph structure. 

Here, we present an energy-based generalization to higher dimensions, inspired by a spectral result from the repelling Laplacian on n-complete graphs. By re-writing the Hamiltonian in Eq. \ref{springham} as a function of distance, we (1) show that $n$-complete negative graphs require an $n-1$-dimensional embedding, given by the first $n-1$ eigenvectors, and (2) we argue that for non-complete graphs, we can find the best dimension by locating the minimum energy of the distance Hamiltonian as a function of dimension.

\subsection{Results on Complete Graphs}
The inspiration for the higher dimensional generalized Hamiltonian comes from the spectrum of the $n$-complete negative graph, which is trivially cluster-able into $n$ antagonistic factions. Intuitively, an optimal embedding should place these $n$ nodes at equidistant positions, due to the equal repulsive forces that act on each of them in the Hamiltonian from Eq. \ref{springham}. These positions can be represented by a (squared) distance matrix $D$ with entries: \begin{equation}
\label{distmat}
   D_{ij} = a, \forall i \neq j \in \mathcal{V}
\end{equation}
Where $D_{ij} = |\textbf{x}_{i} - \textbf{x}_{j}|^{2}$, and $a \not\to \infty$ due to the spherical constraint. Arranging $n$ equidistant points in Euclidean space requires $n - 1$ dimensions. We can indeed recover this $n -1$ dimensional embedding using the repelling Laplacian as follows. The repelling Laplacian corresponding to the $n$-complete negative graph is of the form: 
\begin{equation*}
L_{r} = 
\begin{pmatrix}
-(n-1) & 1 & \cdots & 1 \\
1 & -(n-1) & \cdots & 1 \\
\vdots  & \vdots  & \ddots & \vdots  \\
1 & 1 & \cdots & -(n-1) 
\end{pmatrix}
\end{equation*}
Using the fact that $L_{r} = -nI + J$, where $J$ is the matrix of all 1s, we find that the eigenvalues of $L_{r}$ are 
\begin{align}
\lambda_{n} = \lambda_{n-1} .... = \lambda_{2} = - n \\
\lambda_{1} = 0  
\end{align}
The $n-1$ smallest eigenvalues correspond to $n-1$ orthonormal eigenvectors $\nu_{n}...\nu_{2}$, which describe configurations with equivalent energy for the Hamiltonian in \ref{replapmin}. 

We define $n$ points in $n-1$ dimensions using $\textbf{a}_{i} = (\nu_{n}^{i}, \nu_{n-1}^{i}, ... \nu_{2}^{i})$. The set of points $\textbf{a}_{i}$ define an $(n-1)$-polytope, where each point is equidistant from the others. To see this, note that the matrix defined $A = \frac{-1}{n}L_{r}$ has eigenvalues 1 with multiplicity $n-1$ and eigenvalue 0 with multiplicty 1. It is also symmetric with the same eigenvectors as $L_{r}$, and entries of the form: 
\begin{align*}
A_{ii} &= \frac{n-1}{n}   \\
A_{ij} &= \frac{-1}{n}, \forall i \neq j
\end{align*}
Therefore, it admits a singular value decomposition using the non-zero eigenvectors: 
\begin{equation}
    A = UU^{T}
\end{equation}
where $U = (\nu_{n}, \nu_{n-1}...\nu_{2})$. The entries of matrix $A$ are equivalent to the inner products of the position vectors $\textbf{a}_{i}$, or $A_{ij} =  \textbf{a}_{i} \cdot \textbf{a}_{j}$. Since the relationship between the dot product and the Euclidean distance is given by: 
\begin{equation}
|\textbf{a}_{i} - \textbf{a}_{j}|^{2} = \textbf{a}_{i} \cdot \textbf{a}_{i} + \textbf{a}_{j} \cdot \textbf{a}_{j} - 2(\textbf{a}_{i} \cdot \textbf{a}_{j})
\end{equation}
It follows that $|\textbf{a}_{i} - \textbf{a}_{j}|^{2} = 2$ for all $i \neq j$. We have shown that the multiplicity of the smallest eigenvalue of the repelling Laplacian for an $n$-complete negative graph gives the coordinates for embedding the graph in $n-1$ dimensions. 
\subsection{Higher Dimension Generalized Hamiltonian}
Next, we re-define the one-dimensional Hamiltonian in \ref{replapmin} to include higher dimensions. We will show that the ground state energy of the negative $n$-complete graph embedding is lowest in $n-1$ dimensions. This is simpler if we re-write the Hamiltonian in Eq. \ref{springham} as a function of the distance matrix. Arranging the eigenvectors in order of ascending eigenvalue $\{\nu_{n}, \nu_{n-1},...\nu_{2}, \nu_{1}\}$. Let the distance matrix $D(k)$ correspond to the distances associated to the positions of the nodes found using the first $k$ eigenvectors, such that: 

\begin{equation}
    D(k)_{ij} = |\textbf{x}_{i} - \textbf{x}_{j}|^{2} = \sum_{l = n}^{n-k} (\sum_{i,j} |\nu^{i}_{l} - \nu^{j}_{l}|^{2})
\end{equation}

Where $\nu_{l}$ is the $l$-th eigenvector of the repelling Laplacian. Then, $\textbf{x}_{i}$ is the $k$-dimensional position vector of the $i$-th node constructed using the first $k$ eigenvectors. The distance matrix $D(k)$ is linear in dimension in the sense that if $D(\nu_{l})_{ij} = |\nu^{i}_{l} - \nu^{j}_{l}|^{2}$ is the distance matrix associated with the $l$-th eigenvector, then: 
\begin{equation}
    D(k) = \sum_{l = n}^{n-k} D(\nu_{l})
\end{equation}
The Hamiltonian in \ref{springham} can be written in terms of the distance matrix $D(k)$: 
\begin{equation}
        H(D(k)) = \sum_{i,j}A^{+}_{ij} D(k)_{ij} + \sum_{i,j}A^{-}_{ij} D(k)_{ij}
\end{equation}
Where in one-dimension, 
\begin{equation}
        H(D(1)) = \sum_{i,j}A^{+}_{ij} D(\nu_{n})_{ij} + \sum_{i,j}A^{-}_{ij} D(\nu_{n})_{ij} = \lambda_{n}
\end{equation}
This Hamiltonian is similarly linear in dimension, as follows: 
\begin{equation}
\label{onedimhigh}
        H(D(k)) = \sum_{l = n}^{n-k} H(D(\nu_{l})) 
\end{equation}
Where $H(D(\nu_{l}))$ is simply $\lambda_{l}$, the eigenvalue associated with the $l$-th eigenvector. Then, we have that: 
\begin{equation}
        H(D(k)) = \sum_{l = n}^{n-k} \lambda_{l}
\end{equation}
As a consequence of linearity, the Hamiltonian as a function of distance trivially decreases with an increase in dimension as long as $\lambda_{k} \leq 0$, since each new dimension allows the nodes to be separated by greater distances. This can also be understood by noticing that the norm of the distance matrix will grow with each added dimension. To compare the resulting energy at different embedding dimensions, a natural choice is to normalize the Hamiltonian by the norm of the distance matrix, dividing by the term $\sqrt{(\sum_{i,j}(D(k)_{ij})^{2})}$.  Finally, we have the higher-dimensional generalized Hamiltonian: 

\begin{equation}
\label{higdim}
    \tilde{H}(D(k)) = \frac{\sum_{l=n}^{n-k}\lambda_{l}}{\sqrt{\sum_{ij}(D(k)_{ij})^{2}}}
\end{equation}

Returning to the case of the $n$-complete negative graph, we will show that the higher dimensional generalized Hamiltonian is minimized in $k = n-1$ dimensions. As all possible edges exist and are negative, 
\begin{equation}
\label{completegraphham}
H(D(k)) = -\sum_{i,j} D(k)_{ij} =  -\sum_{l =n}^{n-k} \sum_{i,j}  D(\nu_{l})_{ij} = \sum_{l =n}^{n-k} \lambda_{l}
\end{equation}
Let us denote $vect(D(k))$ as the vectorized distance matrix corresponding to the first $k$ eigenvectors, without the diagonal terms which are trivially 0, so that it has length $n(n-1)$. For $k \leq n -1 $, we have from Eq. \ref{higdim} that: 
\begin{equation}
\label{highdimcomplete}
    \tilde{H}(D(k)) =  \frac{-kn}{|vect(D(k))|}
\end{equation}
By the Cauchy-Schwarz inequality, 
\begin{equation}
\label{CSEQ}
vect(D(k)) \cdot \textbf{1} \leq |\textbf{1}||vect(D(k))|
\end{equation}
Where $vect(D(k)) \cdot \textbf{1} = \sum_{ij} D(k)_{ij}$ and $|\textbf{1}| = \sqrt{n(n-1)}$. By Eq. \ref{completegraphham}, $\sum_{ij} D(k)_{ij} = kn$. Next, we combine Eq. \ref{highdimcomplete} with Eq.\ref{CSEQ} to see that: 
\begin{equation}
\tilde{H}(D(k)) = -\frac{kn}{|vect(D(k))|} \geq -\sqrt{n(n-1)}
\end{equation}
The Cauchy-Schwarz inequality is a strict inequality, unless one vector is a scalar multiple of the other. If $vect(D(k))$ is a scalar multiple of $\textbf{1}$, this implies that the points are equidistant, which is trivially only possible if $k \geq n -1$ dimensions. Indeed, recall from Eq. \ref{distmat} that in $k = n-1$ dimensions the distance matrix is: 
\begin{equation}
    D(n-1)_{ij} = 2, \forall i \neq j
\end{equation}
Such that $vect(D(n-1)) = 2\textbf{1}$. 
It follows that: 

\[
  \tilde{H}(D(k)) 
  \begin{cases}
    > -\sqrt{n(n-1)}, & \text{for } k < n-1 \\
    = -\sqrt{n(n-1)}, & \text{for } k = n-1 
  \end{cases}
\]
Or, the energy is strictly minimized in $k = n-1$ dimensions, precisely as we needed to show. 
\subsection{Generalizations to Non-Complete Graphs}

Using the intuition from the $n$-complete negative graph, this process can be numerically generalized to non-complete graphs to find the best dimension for representing proximal node relationships. We seek to minimize the higher dimensional generalized Hamiltonian given in Eq. \ref{higdim} as a function of $k$, thus finding the embedding dimension that minimizes the ground state energy. Note that even if $\lambda_{k + 1} < 0$, the addition of the $k + 1$-th dimension may still increase the energy $\tilde{H}(D(k+1)) > \tilde{H}(D(k)$ due to the normalization factor. This is related to the idea proposed in \cite{knyazev2018spectral} of using the $k$ smallest eigenvectors for the embedding, where $k$ is chosen by looking for the largest eigen-gap, but formalizes this argument in term of a physical energy function. 

In general, even if a signed network has $n$ ground truth clusters, the relationship between the required dimension $k$ and $n$-clusterability is complicated. In particular, it is possible to represent the distances between $n$ factions in $k < n -1$ dimensions if the weights of the negative forces between the factions are conducive to a lower dimensional embedding. To see this, suppose that a matrix $D^{*}$ represents the set of distances between nodes that minimizes the Hamiltonian. Let $\hat{D}$ be the matrix obtained by squaring the entries of $D^{*}$. According to the Schoenberg theorem (see \cite{bogomolny2007distance}), these distances have an embedding into Euclidean space if:
\begin{equation}
\label{eucembed}
\forall \xi \in \mathbb{R}^n, \sum_{i = 1}^{N} \xi_{i} = 0 \implies \sum_{i,j}\xi_{i}\hat{D}_{ij}\xi_{j} \leq 0  
\end{equation}
Suppose that $\hat{D}$ satisfies Eq. \ref{eucembed}. Define the $n-1$ square matrix $C^{*}$ by 
\begin{equation}
C^{*}_{ij} = \frac{1}{2}(D^{*_{2}}_{in} + D^{*_{2}}_{jn} - D^{*_{2}}_{ij})
\end{equation}
Then, the minimal dimension required for the Euclidean embedding is given in~\cite{deza1997geometry} as: 
\begin{equation}
    k_{min} = rank(C^{*})
\end{equation}
As a more illustrative example, consider a 3-complete negative graph with nodes $i,j,k$ and edge weights defined such that $|\textbf{x}_{i} - \textbf{x}_{j}| = 2$, $|\textbf{x}_{i} - \textbf{x}_{k}| = 3$ and $|\textbf{x}_{k} - \textbf{x}_{j}| = 5$ are the distances the minimize the Hamiltonian. Even though there are 3 "factions", only one dimension is required for the embedding if node $i$ is placed in between nodes $j$ and $k$. Consequently, we emphasize again that the energy minimization is intended to find the dimension that produces the \textit{best proximity measure}, and not the number of weak balance clusters.  

\begin{figure}
\includegraphics[scale=0.20]{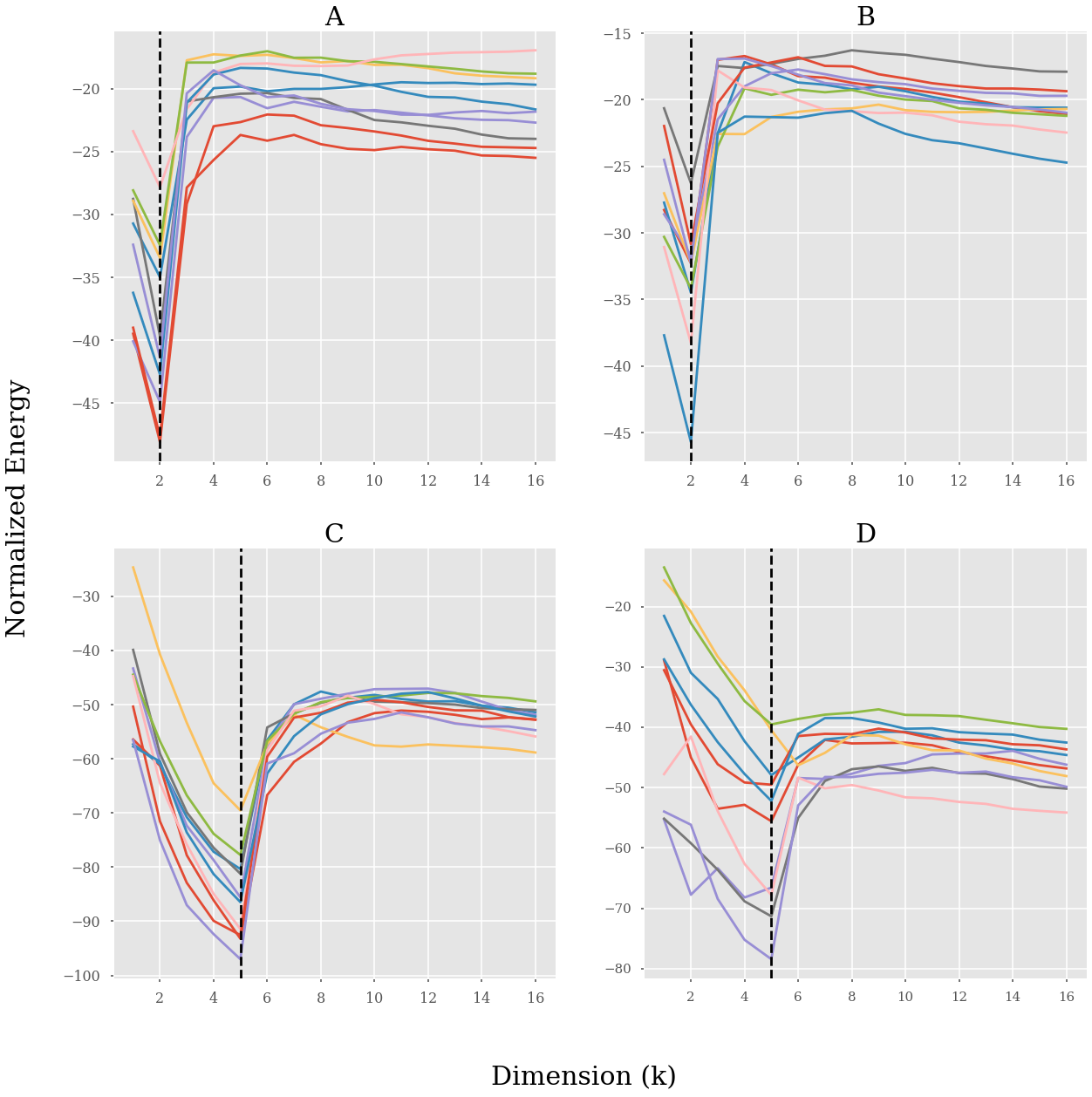}
\caption{Normalized Energy $\tilde{H}(D(k))$ vs dimension $k$ for 10 realizations of different SSBMs. Dashed line shows the energy minima. A) 3 Community SSBM with edge probability 0.5, and randomly generated community sizes between 20 to 50. The energy minimum is clearly obtained at $k = 2$ as expected. B) 3 Community noisy SSBM with edge probability 0.5, and randomly generated community sizes between 20 to 50, with 0.1 probability of sign flip. For these realizations, the energy minimum is still obtained at $k = 2$. C) 6 Community SSBM with edge probability 0.5, and randomly generated community sizes between 20 to 50. The energy minimum is clearly obtained at $k = 5$ as expected. D) 6 Community nosiy SSBM with edge probability 0.5, and randomly generated community sizes between 20 to 50, with 0.1 probability of sign flip. The energy minimum is still obtained  at $k = 5$, for most realizations, but can vary depending on realization faction sizes and random variations. }
\label{fig:weakbalresults}
\end{figure}

\begin{figure}
\centering
\includegraphics[scale=0.20]{sbm100_3com.png}
\includegraphics[scale=0.20]{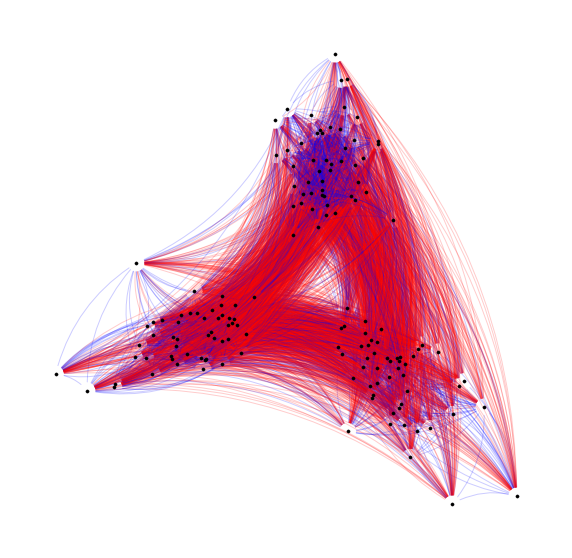}
\caption{3 Community SSBMs with 50 node communities and 0.5 edge probability, embedded using SHEEP (first two eigenvectors of repelling Laplacian). No edge sign flips (left) and 0.2 probability of sign flip (right).}
\label{fig:3com_embed}
\end{figure}

Figure \ref{fig:weakbalresults} shows the normalized energy versus dimension plots for various realizations of SSBMs with different community numbers, sizes and sign flip probabilities. Figure \ref{fig:weakbalresults}A shows the normalized energy as a function of dimension for 3 community SSBM with edge probability 0.5, and randomly generated community sizes between 20 to 50. For each of the 10 realizations, the energy is minimized at $k = 2$, the expected dimension for representing 3 communities. Likewise, in Figure \ref{fig:weakbalresults}B, the energy versus dimension plots for the 3 community SSBMs with noise (0.1 probability of edge sign flip), shows the minimum is still obtained at $k = 2$. Figure \ref{fig:3com_embed} shows a two-dimensional SHEEP embedding for two realizations of a 3 community SSBM, with and without noise on the edge signs.  When the number of SSBM communities is increased to 6 as in Figure \ref{fig:weakbalresults}C, the minimum energy is obtained at dimension $k = 5$ for each of the 10 realizations. For the noisy 6 community SSBMs show in Figure \ref{fig:weakbalresults}D, the energy minimum is obtained at $k = 5$ for most realizations, but can vary depending on the faction size and noise. 

\section{Applications}
\subsection{Proximity Based Node Attributes}
Since our constructed Hamiltonian is a function of distance, SHEEP is designed to represent node proximity, taking into account both local and global interactions. Here, we build on this insight to show how SHEEP can be used to recover continuous node attributes. We emphasize that embedding methods designed for clustering are particularly adapt at predicting binary variables (edge signs), while SHEEP provides a continuous metric for understanding node relationships and relative extremism. In Appendix B, we provide some visual example that illustrate this distinction on various synthetic graphs, comparing the embedding produced by SHEEP with two other spectral methods, the opposing Laplacian and SPONGE~\cite{cucuringu2019sponge}. In the following section, we quantitatively evaluate the performance on both synthetic generated and empirical signed networks. 

\subsubsection{Rankings on Generated Synthetic Graphs}
\begin{figure}
\centering
\includegraphics[scale=0.3]{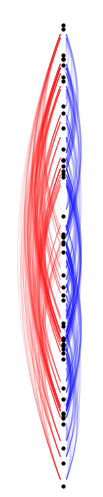}
\includegraphics[scale=0.221]{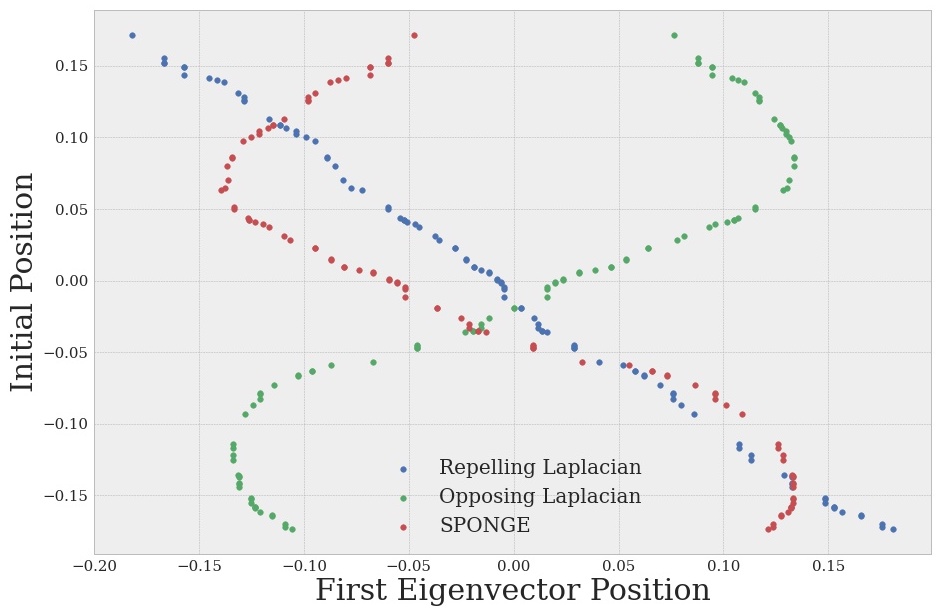}
\caption{Positions of 50 nodes with edge signs assigned according to threshold of 0.2 (left). Generated node position plotted against first eigenvector position from SHEEP (repelling Laplacian), the opposing Laplacian and SPONGE for a 100 node graph with threshold 0.1 (right). There is a high correlation between the initial and SHEEP embedding position.}
\label{fig:Ranking}
\end{figure}

Since the SHEEP Hamiltonian is a function of distance, the sign of an edge between two nodes is related to the distance between them in the resulting embedding. In one dimensions, the positions generated by the SHEEP embedding resembles an ordering of the nodes, where nodes that are closer together in the ordering have more positive connections. In this experiment, we generate a complete signed graph by placing nodes randomly with uniform distribution between -1 and 1, and normalizing the resulting position vector to 1. The signed graph is constructed deterministically: nodes are connected by a positive edge if the distance between them is less than a chosen threshold value, otherwise the edge sign is negative. Figure \ref{fig:Ranking} (left) shows the initial distribution of 50 nodes, with edge signs allocated according to the threshold value of 0.2. We generate the one-dimensional SHEEP embedding using the first eigenvector of the repelling Laplacian, and compare the resulting node ordering to the initial positions by taking the Kendall Tau correlation, denoted as KT. correlation from now on.  Since multiplying the eigenvector by -1 returns the same eigenvalue, but with opposite node ordering, we take the absolute value of the KT. correlation.  We also compare SHEEP to the opposing Laplacian and SPONGE, two spectral methods which were designed for clustering, finding that our method is better at recovering ordinal information.  

\begin{figure}
\centering
\includegraphics[scale=0.25]{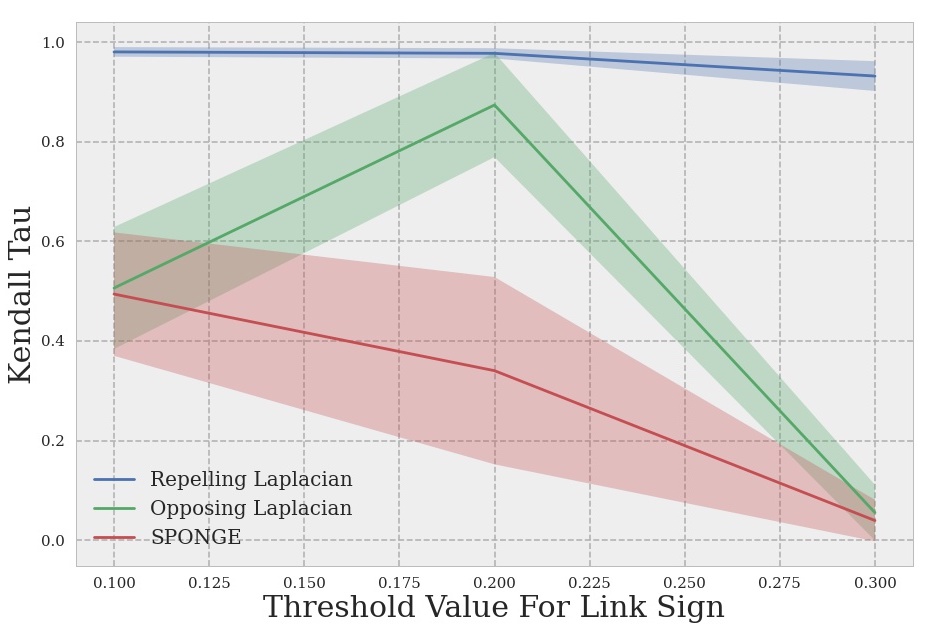}
\caption{N = 50 Nodes, Mean KT. correlation between initial ranking and SHEEP, opposing Laplacian and SPONGE rankings over 100 realizations.  Shaded area gives standard error. SHEEP outperforms the other methods designed for clustering.}
\label{fig:KT50}
\includegraphics[scale=0.25]{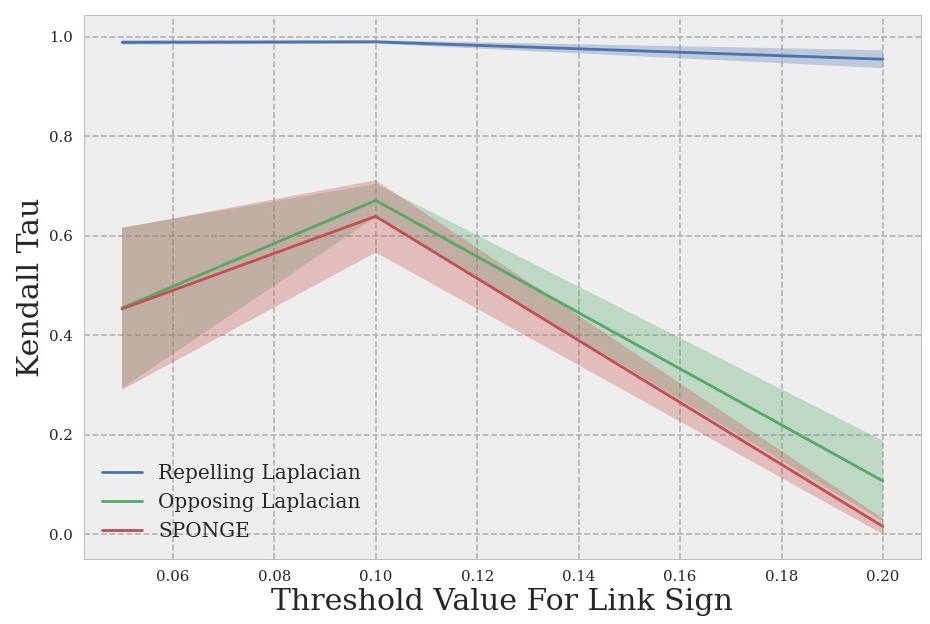}
\caption{N = 100 Nodes, Mean KT. correlation between initial ranking and SHEEP, opposing Laplacian and SPONGE rankings  ranking over 100 realizations. Shaded area gives standard error. Again, SHEEP outperforms the other methods designed for clustering.}
\label{fig:KT}
\end{figure}

Figure \ref{fig:Ranking} (right) shows a plot of the generated node position versus the position determined by the first eigenvector for one realization of the generated graph, with 100 nodes and threshold = 0.1. Unlike the opposing Laplacian and SPONGE, the SHEEP embedding recovers the initial node ordering.  The resulting KT. correlations are displayed for several realizations of initial node distributions and different threshold values in Figures \ref{fig:KT50} and \ref{fig:KT} for $n = 50$ and $n =100$ nodes respectively. For the case of $n = 100$, there is a higher density of nodes, so the threshold values are chosen to be lower. SHEEP is highly successful at recovering the resulting node ordering, unlike the other two spectral methods, demonstrating the usefulness of our proximity-based embedding measure.   

\subsubsection{Australian Rainfall Correlations}
In this section, we test the method on an empirical signed graph with a ground-truth proximity measure. Using a time series of seasonal rainfalls (in mm) across different stations in Australia, we construct a Pearson correlation matrix, resulting in a complete signed graph on $n = 305$ nodes. We mask the edge weight magnitude by taking the sign of the Pearson correlations such that for all $(i,j) \in E$, $(i,j) \in [-1, 1]$. Note that the sign of Pearson correlation is not transitive, depending non-trivially on the magnitude of the correlation. As a result, the resulting signed graph does not have a perfect bi-partition. The same data-set was studied in \cite{cucuringu2019sponge}, where the authors used a $k = 6$ and $k = 10$ dimensional embedding to obtain clusters of stations, which corresponded with the ground truth geographical regions. Using our higher dimensional generalized Hamiltonian method, we find that the best embedding dimension is just $k = 1$. Intuitively we might expect geographically embedded stations would require two dimensions, however, meteorological research has shown that the Australian climate is determined largely by latitude, due to a high-pressure belt which moves north and south over the year, influencing the rainfall patterns over the seasons~\cite{deacon1953climatic}. Since our embedding method can be used to understand proximal relationships and continuous node properties, we can compare the embedding positions to north-south geographic distance. Ideally, the positions generated by our Hamiltonian should be correlated with the latitude of the stations. As in the experiments on the synthetic graphs, we take the absolute value of the KT. correlation between the first repelling Laplacian eigenvector (one-dimensional node positions) and the latitude (which has an approximate linear relationship to north-south distance). Again, we compare to the opposing Laplacian and SPONGE.  

\begin{figure}
\includegraphics[scale=0.25]{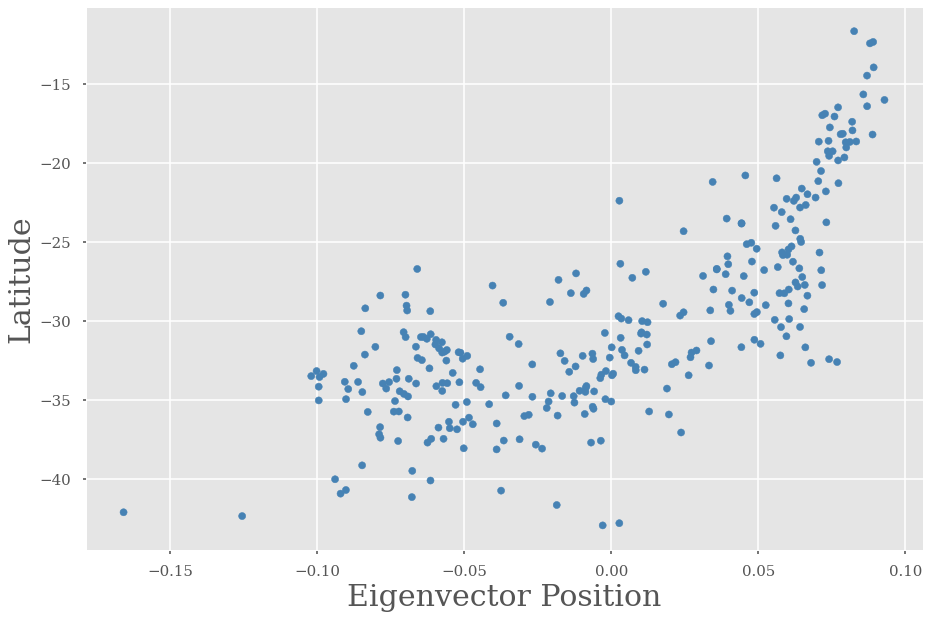}
\caption{Station latitude versus SHEEP embedding positions, obtained from the first eigenvector of the repelling Laplacian. Edge weights are masked using the sign function, $(i,j \in [-1, 1]$. KT. correlation is 0.744. }
\label{fig:ausmasked}
\end{figure}

\begin{table*}[htb]
\label{table}
\caption{KT. Correlation between Eigenvectors of $L_{r}$ (SHEEP), $L_{o}$, SPONGE and Latitude for Australian Rainfall Correlation Network}
\centering
\begin{tabular}{ | m{8em} | m{3em}| m{3em} | m{3em} | m{3em} | m{3em} |} 
\toprule
        & $\lambda^{r}_{n = 305}$ & $\lambda^{o}_{n = 305}$ & $\lambda^{sponge}_{n = 305}$ & $\lambda^{sponge}_{n = 304}$ \\
\midrule
KT. Corr. with Latitude & \textbf{0.744} & 0.719 & 0.178 & 0.733 \\
\bottomrule
\end{tabular}
\end{table*}
As in Table I, the first eigenvector of the repelling Laplacian (SHEEP) has the highest ordinal correlation with station latitude, compared to the first eigenvector of both the opposing Laplacian and SPONGE. We note that the second eigenvector of SPONGE has a higher correlation with latitude as compared to the first. As we will see with the following examples, the \textit{location} of the best SPONGE eigenvector changes for different networks, making it difficult to know which eigenvector to choose without ground truth proximity information. Figure \ref{fig:ausmasked} shows the SHEEP embedding position versus latitude for each station, visually demonstrating the node attribute and embedding relationship.  Even without the edge magnitude information, SHEEP still obtains good correlation with the latitude. 

\subsection{Node Extremism Measure}
\subsubsection{Signed Configuration Model}
In this section, we argue that the distance from the node to the origin in the optimal embedding dimension gives a measure of the node’s “extremism” or conflictuality.  In general, the distance from the node to the origin in the embedding is dependent on both local measures, like the node’s positive and negative degrees, and more global properties like the structure of the graph. For instance, as in Appendix Figure \ref{fig:sbmweakden}, SHEEP places nodes with fewer negative edges to the other factions are closer to the origin. Thus, the distance to the origin is determined by both the cluster structure of the graph, and by the intensity of the negative interactions between the nodes. When the graph structure is random, we expect that local and global measures will coincide. Using a signed graph configuration model, we investigate the relationship between the node’s distance to the origin in the SHEEP embedding, and the net degree (negative degree minus positive degree). The configuration model is a random graph model constructed from a given degree sequence, where each edge “stub” is matched with equal probability~\cite{newman2018networks}. 

\begin{figure}
\includegraphics[scale=0.25]{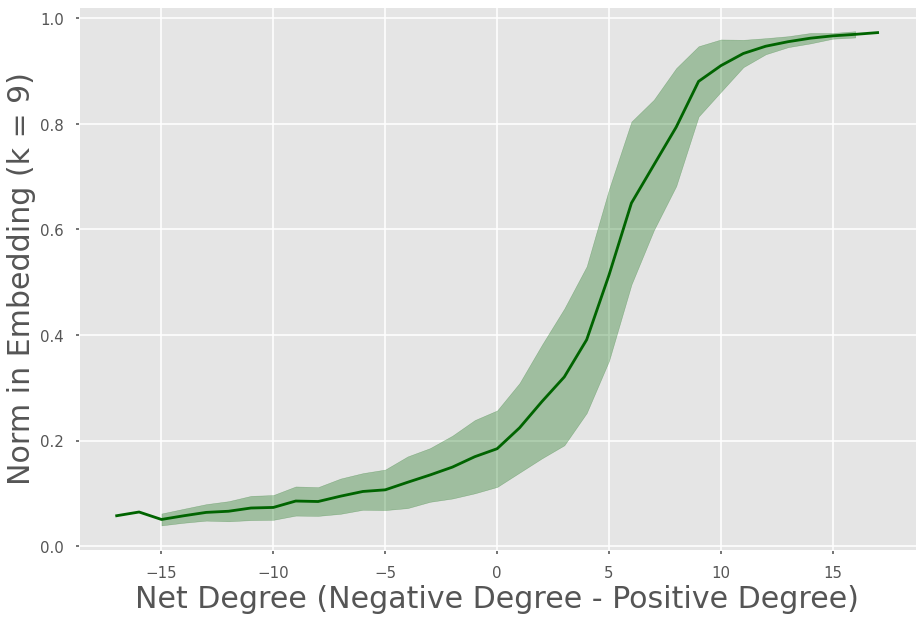}
\caption{Mean distance to origin in SHEEP embedding (the norm in $k = 9$ dimensions) versus the node's net degree over 100 realizations of the signed network configuration model, for which $k = 9$ was the optimal embedding dimension. Shaded area gives the standard deviation. Nodes having relatively more negative edges are further from the origin.}
\label{fig:configsigned}
\end{figure}

For the signed graph configuration model, we generate two 50 node configuration models using randomly generated degree sequences of integers between 1 and 20.  We take the graph obtained by subtracting the adjacency matrix of the first model from the second to obtain a random signed network. Over 5000 realizations of the configuration model, the most frequently obtained optimal embedding dimension is $k=9$ using SHEEP. Using 100 realizations of the configuration models for which $k=9$ was optimal, we find for each node the net degree, and the distance to the origin, which corresponds to the norm of its 9-dimensional position vector. The mean distance for each net degree is shown in Figure \ref{fig:configsigned}, where the shaded area gives the standard deviation over the 100 realizations. We clearly obtain an increasing relationship, where nodes with a higher proportion of negative edges are more “extreme” and thus further from the origin. As in the next section, in the case when the graph is not random and has some structure (eg. factions), the norm still gives a measure of extremism that cannot be directly explained by the net degree. 

\subsubsection{Continuous Political Ideology}

\begin{figure}
\includegraphics[scale=0.24]{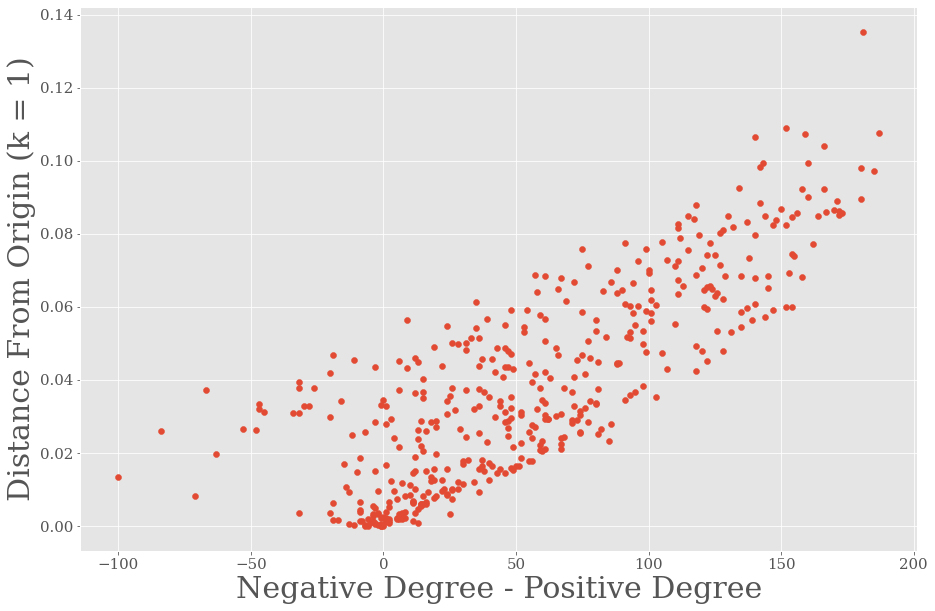}
\caption{Distance from the origin (in one dimensional SHEEP embedding) versus net degree for Congress 114. The KT. correlation is weaker than on the random synthetic graphs (0.614), indicating that graph structure in addition to net degree plays a role in node "extremism".}
\label{fig:NodeExtremeScatter}
\end{figure}

\begin{figure*}
  \includegraphics[width=0.8\textwidth,height=3.9cm]{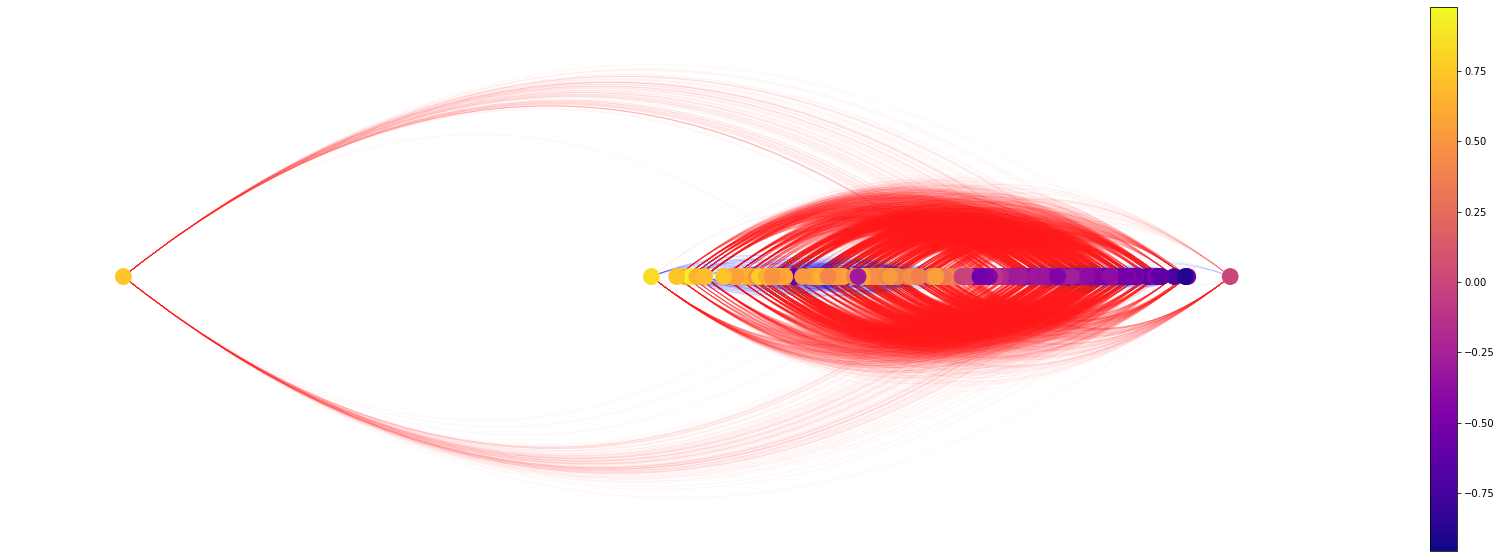}

\includegraphics[width=0.79\textwidth,height=3.9cm]{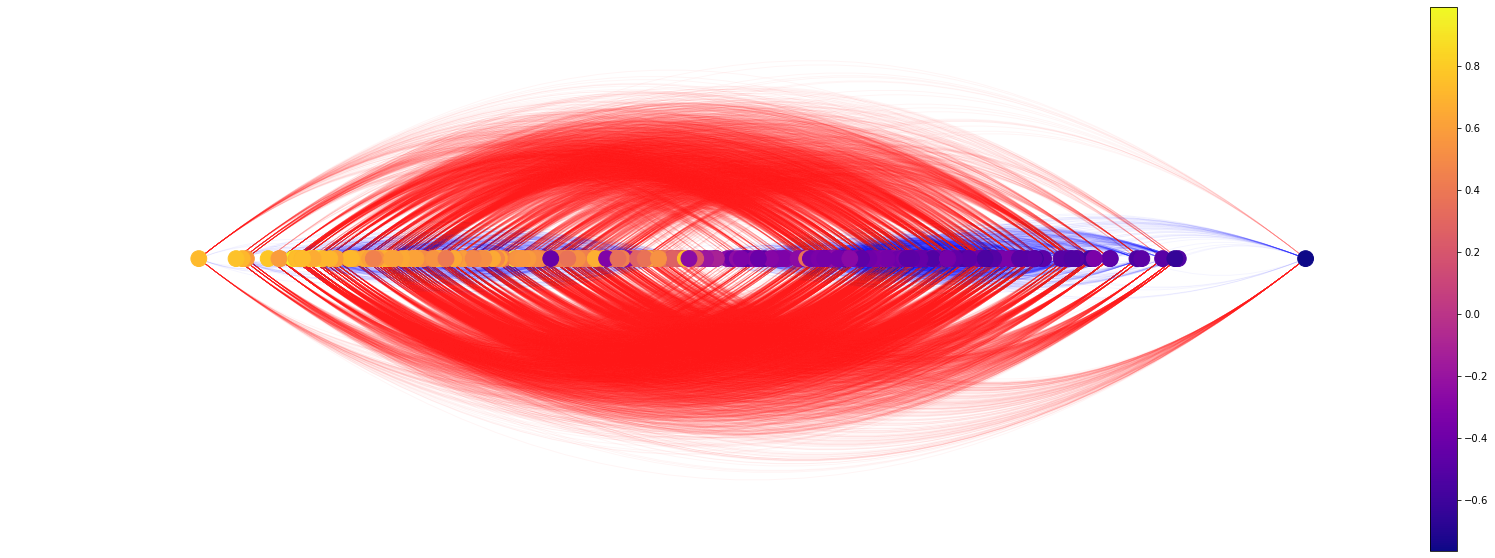}
  
  \caption{One dimensional SHEEP embedding of Congress 110 (top) and Congress 114 (bottom) signed networks. Nodes colours indicate Nokken-Poole political ideology scores obtained from \cite{aref2021identifying}. }
\label{fig:NPSCORE}
\end{figure*}

Here, we consider an empirical network with ground truth continuous node attributes, which provides a better intuition for the node "extremism" measure.  We take a signed network representing relationships between members of the USA House of Representatives for different congresses (here, we focus on sessions 110, and 144). The data-set is obtained from \cite{aref2021identifying}. In this signed network, a positive (negative) edge indicates that the two representatives have co-sponsored statistically significantly more (fewer) bills than expected by chance. Since we focus on the more recent congresses, which Aref et al. argue in \cite{aref2021identifying} are highly bi-polarized along Democrat-Republican lines, it is not surprising that the optimal SHEEP embedding dimension is $k=1$. Because SHEEP represents proximal relationships, we want to understand whether the embedding can recover the political ideology of the members on a continous scale, and investigate whether the distance to the origin in the embedding corresponds with political extremism. As suggested in \cite{aref2021identifying}, we use Nokken-Poole (NP) ideology scores as our ground truth, which place each member of congress on a continuous scale from +1 (conservative) to -1 (liberal). The score is frequently used in political science, and is the result of a three step multi-dimensional scaling method based on member's voting habits in a given congress, derived by maximizing their utility~\cite{carroll2009measuring, nokken2004congressional}. The Nokken-Poole scores are two dimensional, but the first dimension is often taken to represent political ideology on a left/right spectrum.

The 110th congress session forms a signed graph of $n = 452$ nodes. Table II compares the KT. correlations between the positions obtained from SHEEP (first eigenvector of the repelling Laplacian), the opposing Laplacian and SPONGE. The SHEEP embedding obtains the best correlation with the chosen proximity measure for ideology, represented by the Nokken-Poole scores.  In this case, there exists an eigenvector of SPONGE which exhibits a better correlation with the NP scores. Of the first ten smallest eigenvectors, the sixth has the highest KT. correlation with the NP scores. However, as we will see with the second choice of congress, the location of the best SPONGE eigenvector changes as the network changes. Figure \ref{fig:NPSCORE} (top) shows the one-dimensional embedding of the network obtained by SHEEP, where the nodes are colored according to the Nokken-Poole scores. 

\begin{table*}[htb]
\label{table110}
\caption{KT. Correlation between Eigenvectors of $L_{r}$ (SHEEP), $L_{o}$, SPONGE and Nokken-Poole Ideology Scores for USA House of Representatives Network, Congress Session 110}
\centering
\begin{tabular}{ | m{10em} | m{3em}| m{3em} | m{3em} | m{3em} | m{3em} |} 
\toprule
         & $\lambda^{r}_{n = 452}$ & $\lambda^{o}_{n = 452}$ & $\lambda^{sponge}_{n = 452}$ & $\lambda^{sponge}_{n = 447}$ \\
\midrule
KT. Corr. with NP Ideology Score & \textbf{0.684} & 0.003 & 0.005 & 0.639 \\
\bottomrule
\end{tabular}
\end{table*}

The 114th congress session forms a signed graph of $n = 446$ nodes. Table III compares the KT. ordinal correlations for the 114th congress between the positions obtained from SHEEP (the first eigenvector of the repelling Laplacian), the opposing Laplacian and SPONGE. Again, SHEEP obtains the best ordinal correlation with the Nokken-Poole scores. Of the first ten smallest eigenvectors, the third eigenvector of SPONGE has the highest correlation with the NP scores.  As compared to the network of congress session 110, the location of the best proximal SPONGE eigenvector has changed, while for the repelling Laplacian, this information is still contained in the first eigenvector. Knowing \textit{where} this proximal information will occur in the set of SPONGE eigenvectors is non-trivial, especially when the ground truth scores do not exist. For this, SHEEP is superior, as the information is contained in first $k$ eigenvectors deemed optimal by the embedding method.  Figure \ref{fig:NPSCORE} (bottom) shows the one-dimensional SHEEP embedding of the network, where the nodes are once again colored according to the Nokken-Poole scores.

Since the optimal embedding dimension for the 114th congress is $k = 1$, the KT. correlation shows directly the relationship between the node's distance to the origin and its political extremism, using the NP scores. Since this network has structure, unlike in the case of the signed network configuration model, the distance to the origin cannot be directly explained by the net degree. Figure \ref{fig:NodeExtremeScatter} shows the distance to the origin versus the net degree for nodes in the 114th congress network. The KT. correlation is lower than on the random graph experiments, and also lower than the correlation obtained with the NP scores, indicating that the distance to the origin in the SHEEP embedding incorporates more than just local information. 

\begin{table*}[htb]
\label{table114}
\caption{KT. Correlation between Eigenvectors of $L_{r}$ (SHEEP), $L_{o}$, SPONGE and Nokken-Poole Ideology Scores for USA House of Representatives Network, Congress Session 114}
\centering
\begin{tabular}{ | m{10em} | m{3em}| m{3em} | m{3em} | m{3em} | m{3em} |} 
\toprule
         & $\lambda^{r}_{n = 446}$ & $\lambda^{o}_{n = 446}$ & $\lambda^{sponge}_{n = 446}$ & $\lambda^{sponge}_{n = 444}$ \\
\midrule
KT. Corr. with NP Ideology Score & \textbf{0.697} & 0.021 & 0.433 & 0.626 \\
\bottomrule
\end{tabular}
\end{table*}

\section{Conclusion}
In this paper, we have presented SHEEP, a spectral embedding algorithm for finding proximal relationships between nodes in signed networks. The method is based on a physically inspired model: we construct a Hamiltonian that assigns attractive and repulsive forces to the positive and negative edges in the graph. We show that the Hamiltonian is intrinsically related to the graph’s repelling Laplacian, and that finding the minimum energy configuration reduces to an eigenvector problem. Using matrix perturbation theory, we show that the resulting ground state energy, or minimum eigenvalue, is a statistical measure of graph bi-polarization structure. We extend our results to higher dimensions, presenting an energy-based approach to locating the optimal embedding dimension for the network. We propose an application of our measure to recovering proximity based continuous node attributes, showing how the SHEEP embedding reproduces ordinal information on synthetic and empirical networks. We also show that the distance to the origin in the optimal embedding dimension gives a measure of node “extremism”, which is related both to local information like the net degree, and to the graph's global structure. Overall, this work contributes to the growing body of literature on spectral methods for understanding signed networks, and characterizing node relationships by taking into account multi-scale information. 

Future research perspectives include exploring application of SHEEP to signed social media datasets, for a novel interpretation of the node proximity and node extremism measures. The method could also be modified to perform better on sparse networks, perhaps through regularization~\cite{qin2013regularized}. Furthermore, extending the method to be robust to changes in number and density of edges would allow applications to temporal signed networks. In this case, understanding the dynamics of node-to-node proximity could be a rich area for future investigation.

\begin{acknowledgments}
 SB was supported by EPSRC grant EP/W523781/1. The work of RL was supported by EPSRC grants EP/V03474X/1 and EP/V013068/1. We thank the authors of \cite{cucuringu2019sponge} for the access to the Australian Rainfall Correlations data-set.
\end{acknowledgments}

\bibliographystyle{apsrev4-1}
\bibliography{refs}
\clearpage
\onecolumngrid
\section*{Appendix}

\subsection{Repelling Laplacian Spectra Results: Proofs}
\noindent \textbf{Theorem 1.}  \emph{For an n-complete signed graph subject to the condition that $E^{+}, E^{-} \neq \emptyset$, the smallest eigenvalue of the repelling Laplacian is $\lambda_{n} \geq -n.$} \\

\noindent \normalfont \textbf{Proof:} Beginning with the $n$-complete graph of all negative edges, which has the associated repelling Laplacian of form: 
\begin{equation*}
\hat{L}_{r} = 
\begin{pmatrix}
-(n-1) & 1 & \cdots & 1 \\
1 & -(n-1) & \cdots & 1 \\
\vdots  & \vdots  & \ddots & \vdots  \\
1 & 1 & \cdots & -(n-1) 
\end{pmatrix}
\end{equation*}
With eigenvalues: 
\begin{align}
\lambda_{n} = \lambda_{n-1} .... = \lambda_{2} = - n \\
\lambda_{1} = 0  
\end{align}
Any other graph in the set $\{G^{n}\}$ can be reached by successively changing the sign of an edge $(i,j)$ from negative to positive.  Suppose we choose graph $G^{n}_{1}$ which has one positive edge. Without loss of generality, we can assume this edge occurs between nodes $i = 1$ and $j = 2$.  For any other edge $(i,j)$, the proof below holds up to multiplication on the left and right by a permutation matrix $P$, which does not change the resulting eigenvalues, a consequence of the orthogonal decomposition of the repelling Laplacian given in Eq. \ref{orthodecomp}.  
\begin{equation}
\label{shiftlap}
L_{r}^{G^{n}_{1}}  = \hat{L}_{r} + L^{shift}
\end{equation}
Where: 
\begin{equation*}
L^{shift} = 
\begin{pmatrix}
2 & -2 & 0 & \cdots & 0 \\
-2 & 2 & 0 & \cdots & 0 \\
0 & 0 & 0 & \cdots & 0 \\
\vdots  & \vdots  & \ddots & \vdots  \\
0 & 0 & 0 & \cdots & 0
\end{pmatrix}
\end{equation*}
The matrix $L^{shift}$ is clearly positive semi-definite with eigenvalue 0 of multiplicity $n-1$ and eigenvalue $\lambda_{1}(L^{shift}) = 4$. Since all matrices involved in Eq. \ref{shiftlap} are symmetric and real, and therefore Hermitian, we can apply Weyl's inequality on the sum of two matrices to find that: 
\begin{equation}
    \lambda_{n}(\hat{L}_{r}) + 0 \leq \lambda_{n}(L_{r}^{G^{n}_{1}})
\end{equation}
Or equivalently, 
\begin{equation}
    -n \leq \lambda_{n}(L_{r}^{G^{n}_{1}})
\end{equation}
Where we recall that $\lambda_{n}$ is the smallest eigenvalue of the matrix. This argument can be inductively applied to any graph in the set  $\{G^{n}\}$ with more than one positive edge, directly implying that the ground state energy of the $n$-complete  signed graph is bounded from below by $-n$.
\\ \\
\noindent \textbf{Theorem 2.}  \emph{Consider a strongly balanced graph $\tilde{G} \in \{G^{n}\}$, such that the nodes of graph $\tilde{G}$ can be partitioned into two sets $V_{1}$ and $V_{2}$, where $V_{1}, V_{2} \neq \emptyset$. Nodes inside each set are connected with positive edges, while the edges connecting the sets are negative. The smallest eigenvalue of the repelling Laplacian is $\lambda_{n} = -n.$} \\

\noindent \normalfont \textbf{Proof:} Since $\tilde{G}$ is a complete graph, there are $V_{1}V_{2}$ negative edges. We want to find the eigenvector $\nu$ associated to the minimal eigenvalue of $L^{\tilde{G}}_{r}$, which we call $\lambda_{n}$. We also know that the Laplacian $L^{\tilde{G}}_{r}$ has a 0 eigenvalue corresponding to the vector $\frac{1}{\sqrt{n}}\textbf{1}$. Since the eigenvectors of $L^{\tilde{G}}_{r}$ are orthogonal, and assuming $\lambda < 0$, then we have the two conditions that:
\begin{align}
\sum_{i} \nu^{2}_{i} &= 1 \\
\sum_{i} \nu_{i} &= 0   
\end{align}
In order to minimize the energy associated to the positive edges, set $\nu_{i} = x_{1}, \forall i \in V_{1}$ and $\nu_{i} = x_{2}, \forall i \in V_{2}$. From the above conditions, we have that 
\begin{align}
\label{cond1}
V_{1}(x_{1})^{2} + V_{2}(x_{2})^{2} &= 1 \\
\label{cond2}
V_{1}x_{1} + V_{2}x_{2} &= 0
\end{align}
The ground state energy: 
\begin{equation}
    E^{0}_{bal} = \nu^{T}L^{\tilde{G}}_{r}\nu = -V_{1}V_{2}(x_{1} - x_{2})^{2} = \lambda_{n}
\end{equation}
Using conditions given in Eqs. \ref{cond1} and \ref{cond2}, along with the fact that $V_{1} + V_{2} = n$, we have that: 
\begin{equation}
    E^{0}_{bal} = \lambda_{n}= -n
\end{equation}
By Theorem 1, this is the minimum eigenvalue. 
\\ \\ 
\noindent \textbf{Theorem 3.}  \emph{Consider a strongly balanced graph $\tilde{G} \in \{G^{n}\}$, such that the nodes of graph $\tilde{G}$ can be partitioned into two sets $V_{1}$ and $V_{2}$, where $V_{1}, V_{2} \neq \emptyset$. Introduce frustration into the graph by switching two edge signs. When the graph is sufficiently large such that $n > 4$, and $V_{1}, V_{2} > 2$, the ground state energy is strictly increased $E^{0}_{frus} > E^{0}_{bal}$.} \\

\noindent \normalfont \textbf{Proof:} Without loss of generality, assume that $|V_{1}| \geq  |V_{2}|$. Next, we introduce a perturbation to the Hamiltonian associated with the switching of signs of a positive and negative edge. Assume that nodes $i = 0, 1, 2 \in V_{1}$ and $i = 3 \in V_{2}$, and we switch the signs of the edges $(0,1) = 1$ and $(2,3) = -1$. Then this "switching" perturbation is associated with the matrix: 
\begin{equation*}
L^{switch} = 
\begin{pmatrix}
-2 & 2 & 0 & 0 & \cdots & 0 \\
2 & -2 & 0 & 0 & \cdots & 0 \\
0 & 0 & 2 & -2 & \cdots & 0 \\
0 & 0 & -2 & 2 & \cdots & 0 \\
\vdots  & \vdots  & \ddots & \vdots  \\
0 & 0 & 0 & 0 & \cdots & 0
\end{pmatrix}
\end{equation*}
We can easily calculate that $L^{switch}$ has eigenvalues -4, 4, and 0 with multiplicity $n-2$. We take $\mu_{n}, \mu_{n-1}...\mu_{1}$ to be the associated eigenvectors, ordered according to the increasing eigenvalue. These vectors span $\mathbb{R}^{n}$ because $L^{switch}$ is symmetric. Note that $\mu_{n} = \frac{1}{\sqrt{2}}(1, -1, 0,0...0)$ is associated with eigenvalue -4. Finally, the Laplacian of the frustrated graph is: 
\begin{equation}
\label{fruslap}
L_{r} = L^{\tilde{G}}_{r} + L^{switch}   
\end{equation}
We call this graph "frustrated" because the two terms $L^{\tilde{G}}_{r}$ and $L^{switch}$ cannot be simultaneously minimized. To see this, take $\nu_{n}$ to be the eigenvector associated with the minimum in Eq. \ref{gse} for the balanced graph, which has the form $\nu_{n} = (x_{1}, x_{1}, x_{1}, x_{2},....)$, where, as before, each node is assigned position $x_{1}$ or $x_{2}$ according to its faction identity. Clearly, $\nu_{n}$ and $\mu_{n}$ are orthogonal. Furthermore, $L^{switch}$ contributes an increase in energy of $\nu_{n}^{T}L^{switch}\nu_{n} = +2(x_{1} - x_{2})^{2}$.  We want to show that the ground state energy of the frustrated Laplacian is larger than $-n$. This requires the following three results, the first is a result of a theorem proved in \cite{akbari2018spectrum}: 

\begin{enumerate}
   \item The spectrum of the opposing Laplacian of a balanced graph is equivalent to the spectrum of the unsigned graph.
   \item The spectrum of an unsigned complete graph has smallest eigenvalue of 0, and all other eigenvalues are $n$, with multiplicity $n-1$
   \item $L_{o} = L_{r} + 2D^{-}$
\end{enumerate}

\noindent It follows that the second smallest eigenvalue of the opposing Laplacian for the balanced graph is $n$. The maximum possible entry of $2D^{-}$ is $2V_{1}$, if a node in $V_{2}$ is connected negatively to all members of the other faction, $V_{1}$.  Applying Weyl's inequality on the relation given in result 3, to find a bound on the second smallest eigenvalue of the repelling Laplacian of the balanced graph $\lambda_{n-1}$: 
\begin{equation}
\label{bound}
\lambda_{n-1} \geq n - 2V_{1}  
\end{equation}
Recall that $\lambda_{n} = -n$ for the balanced graph. We use Eq. \ref{bound} to give an approximation for the "energy gap" for the balanced graph, between the ground and first energy levels, and we define $\nabla E = \lambda_{n-1} - \lambda_{n} \geq 2(n - V_{1})$.
\\ \\
Next, let $\alpha_{n}$ be the eigenvector associated to the ground state of the frustrated Laplacian in Eq.\ref{fruslap}. There is a decomposition of $\alpha_{n}$ in terms of the eigenvectors of the balanced Laplacian which takes the form: 
\begin{equation}
\alpha_{n} = \sum^{n}_{i = 1}  c_{i}\nu_{i}
\end{equation}
We can also use the fact that $\nu_{n}$ and $\mu_{n}$ are orthogonal to define a new orthonormal basis of $\mathbb{R}^{n}$. This new orthonormal basis $\{x_{n}, x_{n-1}...x_{1}\}$ has $x_{1} = \nu_{n}$ and $x_{n} = \mu_{n}$. We can now also express $\alpha_{n}$ in terms of this new basis:
\begin{equation}
\alpha_{n} = \sum^{n}_{i = 1}  d_{i}x_{i}
\end{equation}
Using the orthogonality of the basis vectors, it follows that $d_{1} = c_{n}$. From this, we have the following two decompositions: 

\begin{align}
\label{ballapdecomp}
\alpha_{n} &= c_{n}\nu_{n} + \sum^{n - 1}_{i = 1}  c_{i}\nu_{i} \\
\label{switchlapdecomp}
&= c_{n}\nu_{n} + \sum^{n - 1}_{i = 2}  d_{i}x_{i} + d_{n}\mu_{n}
\end{align}

\noindent For notational purposes, set $\alpha_{n}' = \alpha_{n} - d_{n}\mu_{n}$. The ground state energy associated with the frustrated Laplacian in Eq. \ref{fruslap} is:
\begin{equation}
E^{0}_{frus} = \alpha_{n}^{T}L^{\tilde{G}}_{r}\alpha_{n} + \alpha_{n}^{T}L^{switch}\alpha_{n}
\end{equation}
Using Eq. \ref{ballapdecomp} to simplify the first term on the right hand side, and equation Eq. \ref{switchlapdecomp} to simplify the second term: 
\begin{equation}
\label{efrust}
E^{0}_{frus} = \sum^{n}_{i = 1}  c^{2}_{i}\lambda_{i} + d_{n}^{2}(-4) + \alpha_{n}{'^T}L^{switch}\alpha'_{n}
\end{equation}
Because $\alpha_{n}'$ is orthogonal to $\mu_{n}$ by construction, it has a decomposition in terms of $\mu_{n-1}, \mu_{n-2}...\mu_{1}$, which are eigenvectors of $L^{switch}$ with eigenvalues greater than or equal to 0.  Since the third term in Eq. \ref{efrust} must be greater than or equal to 0, $E^{0}_{frus}$ is minimized when $d_{n}^{2}$ is maximized. From \ref{switchlapdecomp}, $d_{n}^{2} \leq 1 - c_{n}^{2}$. 

\begin{equation}
E^{0}_{frus} \geq \sum^{n}_{i = 1}  c^{2}_{i}\lambda_{i} -4(1 - c_{n}^{2})
\end{equation}

\noindent We can now use the energy gap to bound the ground state energy from below: 

\begin{equation}
E^{0}_{frus} \geq 
c^{2}_{n}(-n + 4) + \sum^{n -1}_{i = 1}  c^{2}_{i}(-n + \nabla E) -4
\end{equation}
Imposing the conditions that $n > 4$ and $V_{1}, V_{2} > 2$, such that each faction has more than two members. Then, $\nabla E > 4$, and since $\sum_{i = 1}^{n}c_{i}^{2} = 1$, the strict inequality follows: 
\begin{equation}
E^{0}_{frus} > -n  = E^{0}_{bal}
\end{equation}

\subsection{Clustering vs Proximity: Illustrative Examples}
\subsubsection{Complete Strong Balanced Graph (Bi-Partition)}
\begin{figure}
\centering
\includegraphics[scale=0.4]{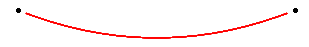}
\includegraphics[scale=0.4]{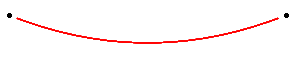}
\includegraphics[scale=0.4]{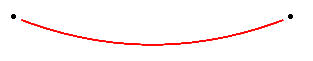}
\caption{Resulting embedding of a 100 node complete graph with bi-partition, which obeys strong balance. From left to right: SHEEP (Repelling Laplacian), opposing Laplacian, SPONGE}
\label{fig:completebi}
\end{figure}

We begin with an $n = 100$ node complete graph with a perfect bi-partition, such that the graph also obeys strong balance. As a result of the bi-partition, only one dimension is needed to represent the graphs. As in Figure \ref{fig:completebi}, each of the embedding methods produce the exact same embedding, as expected for the most trivial case. 

\subsubsection{Non-Complete Strong Balanced Graph (Bi-Partition)}

\begin{figure}
\centering
\includegraphics[scale=0.35]{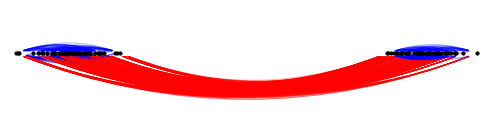}
\includegraphics[scale=0.5]{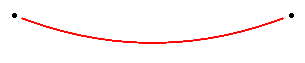}
\includegraphics[scale=0.35]{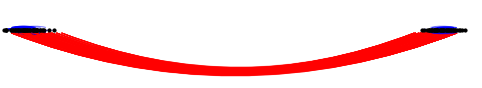}
\caption{Resulting embedding of a [50,50] signed stochastic block model with bi-partition, which obeys strong balance. From left to right: SHEEP (Repelling Laplacian), opposing Laplacian, SPONGE}
\label{fig:sbmbi}
\end{figure}

Here, we generalize to a non-complete graph, using a signed stochastic block model with two communities of size 50, and an affinity matrix of form [[0.5, 0.5], [0.5, 0.5]], where the inter (intra) community edges are negative (positive). Because there is a bi-partition, this graph also obeys strong balance. As in Figure \ref{fig:sbmbi}, the embeddings of SHEEP and SPONGE are once again quite similar. The opposing Laplacian produces a bi-partition where the nodes have been assigned a location in [-1, 1], depending on their cluster assignment. 

\subsubsection{Non-Complete Strong Balanced Graph: Edge Density}

\begin{figure}
\centering
\includegraphics[scale=0.33]{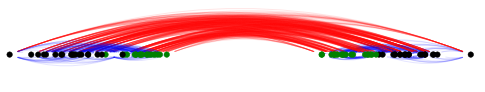}
\includegraphics[scale=0.33]{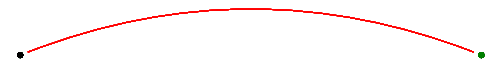}
\includegraphics[scale=0.33]{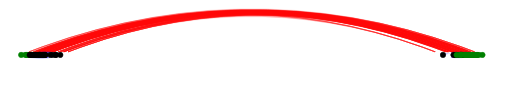}
\caption{Resulting embedding of grouped signed stochastic block model with bi-partition, which obeys strong balance. The green nodes have fewer negative edges compared to the black nodes. From left to right: SHEEP (Repelling Laplacian), opposing Laplacian, SPONGE}
\label{fig:sbmbiden}
\end{figure}

To investigate the relationship between embedding distance and node relationships, we introduce a modified signed stochastic block model which obeys strong balance. Each signed community in the SSBM is separated into two groups of nodes: the first group is connected to the other community with 0.7 density of edges (all negative), and the second group is connected to the other with a 0.3 density of edges (all negative).  Essentially, the two signed communities have been split into two groups, with different probabilities of negative edge existence. Because there is still a bi-partition, this graph also obeys strong balance. In Figure \ref{fig:sbmbiden}, the nodes in green have \textit{lower} densities of negative edges, while the black nodes have \textit{higher} densities of negative edges. As in Figure \ref{fig:completebi}, the opposing Laplacian finds the same partition as in the case of the complete graph. In the embedding produced by SHEEP, the green nodes, which are the nodes with fewer negative edges, are closer to the origin, compared to the black nodes, giving better intuition for the node "extremism" measure. This indicates that the repelling Laplacian embedding distance is better at representing the strength of node relationships, taking into account the presence (or absence) of edges, as compared to both SPONGE and the opposing Laplacian. 

\subsubsection{Weak Balanced Graph: Edge Density}

\begin{figure}
\centering
\includegraphics[scale=0.24]{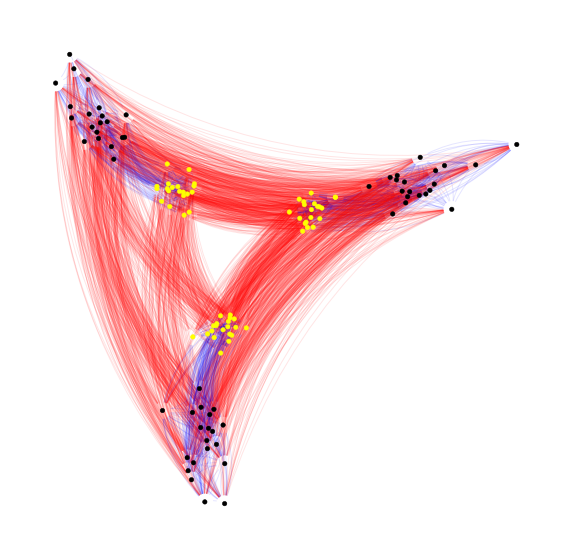}
\includegraphics[scale=0.24]{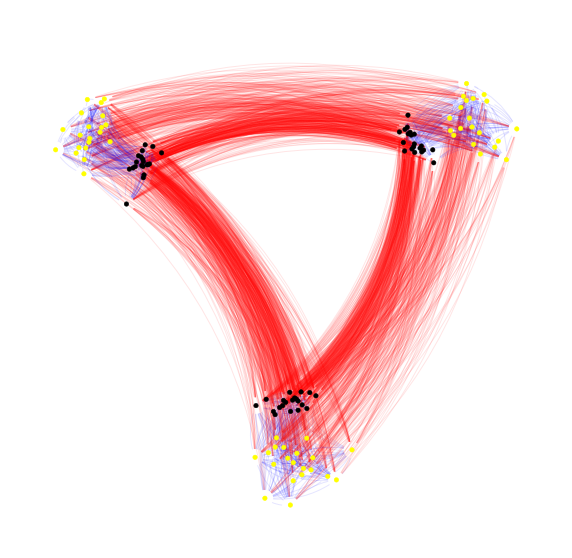}
\includegraphics[scale=0.24]{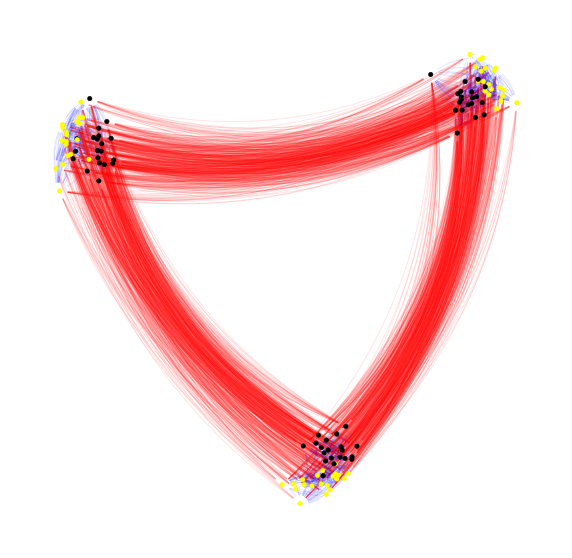}
\caption{Resulting embedding of grouped signed stochastic block model with 3-way partition, which obeys weak balance. The yellow nodes have fewer negative edges compared to the black nodes. From left to right: SHEEP (Repelling Laplacian), opposing Laplacian, SPONGE}
\label{fig:sbmweakden}
\end{figure}

Here, we generalize the ideas above to the case of weak balance, using a 3 community signed stochastic block model with 2 groups in each community in Figure \ref{fig:sbmweakden}. The yellow nodes have a lower density of negative edges to the other communities, as compared to the black nodes. Only in the embedding produced by SHEEP do the yellow nodes appear closer to the origin, indicating the weaker negative relationships between these groups, as compared to the black nodes.

\subsubsection{Weak Balanced Graph: Edge Signs}
\begin{figure}
\centering
\includegraphics[scale=0.24]{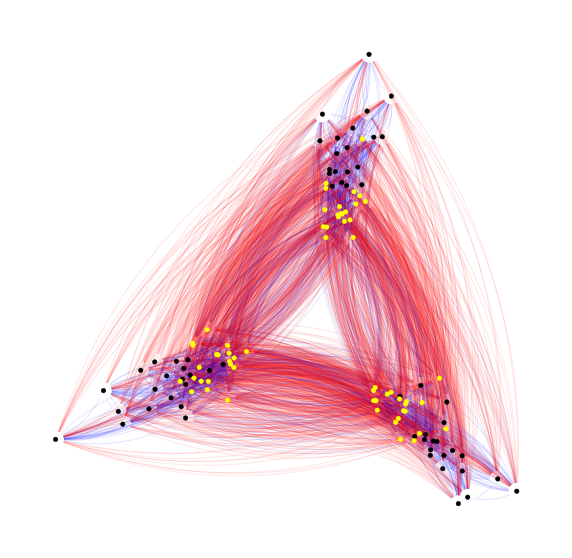}
\includegraphics[scale=0.24]{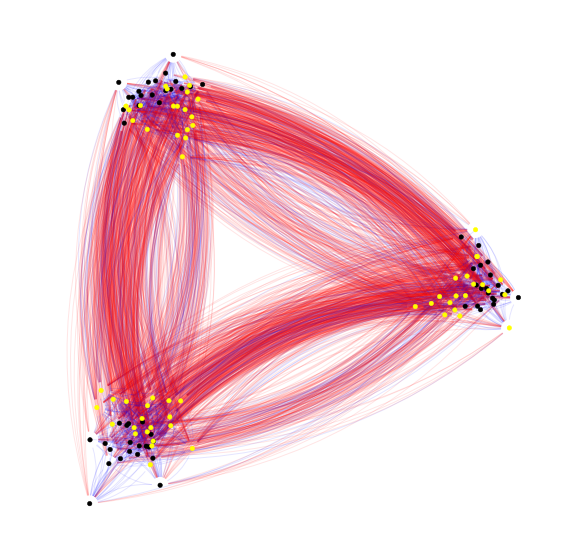}
\includegraphics[scale=0.24]{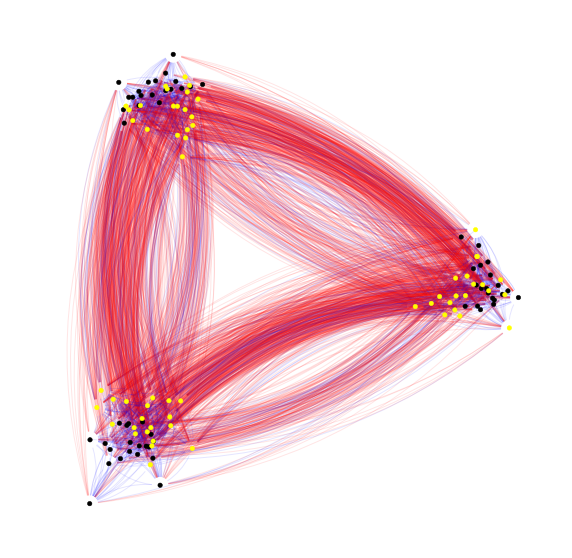}
\caption{Resulting embedding of grouped signed stochastic block model with 3-way partition, which obeys weak balance. The yellow nodes have a 0.2 percent probability of having positive edges to opposing factions, as compared to black nodes which have only negative edges to opposing factions. From left to right: SHEEP (Repelling Laplacian), opposing Laplacian, SPONGE}
\label{fig:sbmwedgesignflip}
\end{figure}

In the final example in Figure \ref{fig:sbmwedgesignflip}, we choose a network that requires $k = 2$ dimensions, obeying weak balance.  We use a 3 community SSBM with 2 groups in each community. Instead of changing the \textit{density} of negative edges, we introduce \textit{noise} on the edge signs on some of the nodes in each signed community. The first group of nodes (black) have the correct sign with 100 percent probability, while the second group of nodes (yellow) have a 20 percent probability of having positive connections to the other factions. Again, only in the embedding produced by SHEEP do the yellow nodes appear in closer proximity, and closer to the origin, indicating the weaker negative relationships between these groups, as compared to the black nodes. The embedding produced by SHEEP gives proximal information that identifies more "extreme" or "neutral" members of each faction and gives a direct relationship between embedding distance and node relationship.

\end{document}